\documentclass[aps,pre,reprint,twocolumn,superscriptaddress,10pt]{revtex4-2}
\usepackage{amssymb,amsmath}
\usepackage[table]{xcolor}
\usepackage{graphicx}
\usepackage{mathtools}
\usepackage[export]{adjustbox}
\usepackage{overpic}
\usepackage[colorlinks=true,
 linkcolor=black,
 urlcolor=blue,
 citecolor=blue]{hyperref}
\usepackage{nicefrac}
\usepackage{multirow}
\usepackage{lipsum}
\usepackage[normalem]{ulem}
\usepackage{pbox}
\newcolumntype{C}[1]{>{\centering\let\newline\\\arraybackslash\hspace{0pt}}m{#1}}
\usepackage{verbatim}
\usepackage{enumitem}

\usepackage{color}
\usepackage{epsfig}
\definecolor{shadecolor}{rgb}{0.85,0.80,0.80}

\definecolor{myorange}{RGB}{253, 184, 99}
\definecolor{mypurple}{RGB}{178, 171, 210}

\newcommand{\comments}[1]{}
\usepackage{multirow}

\usepackage{float}

\newcommand{\avg}[1]{\left\langle{#1}\right\rangle}

\newcommand{\beq}{\begin{equation}}
\newcommand{\eeq}{\end{equation}}
\newcommand{\bal}{\begin{aligned}}
\newcommand{\eal}{\end{aligned}}

\newcommand{\be}{\begin{equation}}
\newcommand{\ee}{\end{equation}}
\newcommand{\bd}{\begin{displaymath}}
\newcommand{\ed}{\end{displaymath}}
\newcommand{\BE}{\begin{eqnarray}}
\newcommand{\EE}{\end{eqnarray}}

\newcommand{\bx}{\ensuremath{\mathbf{x}}}

\allowdisplaybreaks

\begin{document}
\title{A diffusion approximation for systems with frequent weak resetting}

\author{Tobias Galla}
\email{tobias.galla@ifisc.uib-csic.es}
\affiliation{Instituto de F{\' i}sica Interdisciplinar y Sistemas Complejos IFISC (CSIC-UIB), 07122 Palma de Mallorca, Spain}

\begin{abstract}
We develop a diffusion approximation for systems subject to fast random resetting by small amplitudes. Equivalently, this describes systems with frequent but small catastrophes. We demonstrate the validity of the approximation by computing the stationary distribution and mean first-passage times of simple one-dimensional systems. The approximation captures dynamically induced correlations in multi-particle systems, and it can be used to generalise the conditionally independent and identically distributed structure recently found in systems with full resetting. Finally, we show that resetting can induce cycles and patterns, which can be characterised using the diffusion approximation.
\end{abstract}

\maketitle

Even though the theory of stochastic processes has been used in statistical physics for more than a century, new fundamental questions and applications continue to be unearthed. One of these topics centres around systems with resetting. Introduced 15 years ago by Evans and Majumdar \cite{evans2011diffusion} this concerns processes that follow an internal dynamics most of the time, but are `reset' at random times. In the simplest scenario, resetting results in a return to a fixed position.

An impressive analytical apparatus has been developed for systems with resetting, covering Poissonian and non-Poissonian resetting, search problems, extended systems, and quantum dynamics. Reviews can be found in \cite{evans2020stochastic, Gupta}. Recent experimental realisations of resetting use colloidal particles in harmonic traps \cite{tal2020experimental, besga, faisant, biroli2025experimental}. One particularly interesting aspect is the emergence dynamically induced correlations in multi-particle systems where particles move independently, but are reset simultaneously \cite{biroli2023extreme, biroli2024dynamically}.

Stochastic resetting can be interpreted as a series of discrete shocks, akin to shot noise. Processes with resetting are also close to `impulsive dynamical systems'  \cite{bainov1989systems,li2005switched}. If the coordinate of the dynamics is $x$ and resetting is to the origin, the size of each shock is $-x$. If resetting is only partial as in \cite{tal-friedman2022, biroli2024resetting}, say from $x$ to $ax$ with $|a|<1$, the size of a shock is $-(1-a)x$. 

In population dynamics partial resets can describe a catastrophe \cite{Brockwell_Gani_Resnick_1982}, in which a proportion of a population is eliminated. We can for example imagine a process of the form $\dot x=f(x)$, where $x\geq 0$ is the size of a population. If catastrophes happen with rate $r$ and if in any one reset a fraction $(1-a)$ of the population is removed, then the average change of the population size per unit time due to  resets is $-r(1-a)x$. Naively, we therefore expect that the system with catastrophes is approximated the equation $\dot x = f(x)-r(1-a)x$. This deterministic description neglects the random timing of the resets. 

One main goal of this letter is to go further than this. We develop and test a diffusion approximation for systems with frequent but weak resetting. This means to `smear out' a dense series of small shocks to obtain Gaussian randomness, resulting in an effective stochastic differential equation. We do this for single and multi-particle dynamics, and for individual-based systems. We also show that stochastic resetting can induce quasi-cycles and quasi-patterns. The cycles and patterns are not captured by a naive deterministic description, but can be predicted from the diffusion approximation.

The idea of approximating shot noise with Gaussian noise is well established, see e.g. \cite{rosenbaum2016diffusion,TAMBORRINO2021132845} and references therein. The dynamics of chemical reaction systems and populations for example proceeds by discrete events (e.g., conversion of a molecule, or death of an individual). If the system is large, many events occur per unit time, but the relative effect of each event is small. The so-called `chemical Langevin equation' can then be derived \cite{kurtz1978strong, van1992stochastic, gardiner2004handbook,GillespieCLE}. The noise here originates from the internal dynamics, it is {\em intrinsic}. Instead, the resets (or catastrophes) in this work are events that are {\em extrinsic} to the system proper. As far as we are aware diffusion approximations have not been discussed in this context.
\medskip

{\em Diffusion approximation.} We start by looking at a single variable $x$ undergoing a process of the type
\be\label{eq:1dmodel}
\dot x(t)= f[(x(t)]+\xi(t)
\ee
between resets, where $f(x)$ is some function, and where $\xi(t)$ is Gaussian  noise of mean zero and with $\avg{\xi(t)\xi(t')}=2D\delta(t-t')$. With  rate $r$ the particle is (partially) reset ($x\to ax$), where $a\in[0,1]$ is a model parameter. The choice $a=0$ corresponds full resetting, $a=1$ means no resetting at all. In principle, $a$ can depend on the state of the system just before the reset, i.e., $a=a(x)$.

\begin{figure}
    \centering
      \includegraphics[width=0.95\linewidth]{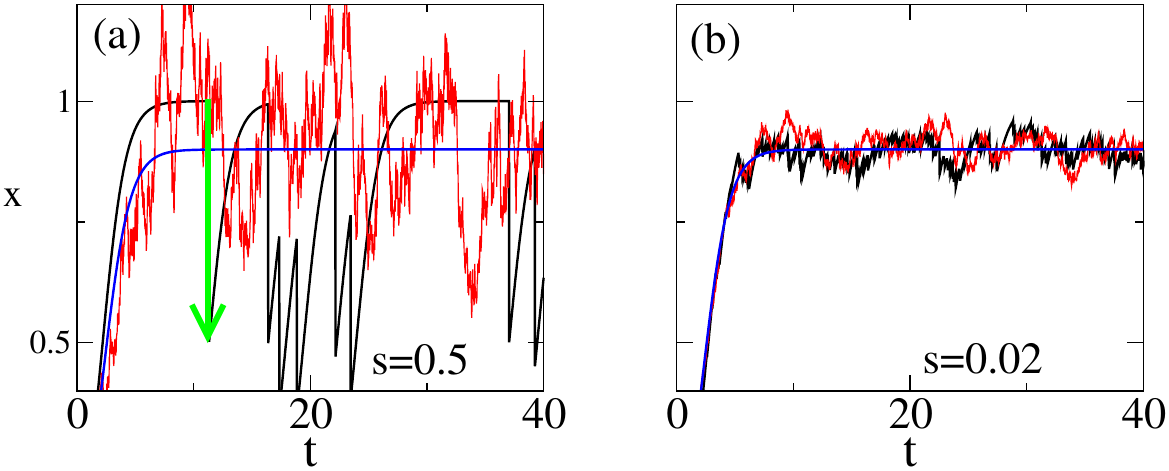}
        \caption{Resetting process and stochastic differential equation for the logistic equation $\dot x=x(1-x)$, with $\lambda=0.1$, $D=0$ and proportional resets $g(x)=x$. Panel (a) $s=0.5$ ($50\%$ removal), panel (b) $s=0.02$. Black curves show a realisation of the system with resetting, green arrow in (a) illustrates one resetting event. Red curves realisations of the SDE (\ref{eq:effective_eq}). Blue curves are the deterministic limit ($s=0$).
    \label{fig:fig1}}
\end{figure}

An illustration is shown in Fig.~\ref{fig:fig1}. We set $D=0$ and use logistic growth, $\dot x=f(x)=x(1-x)$ subject to random resets with rate $r=1/s$, and where a proportion $s$ of the population is removed in each reset. Panel (a) is for $s=0.5$, the discrete resets are clearly visible (black curve). Panel (b) is for $s=0.02$, so that catastrophes are 25 times more frequent than in (a) but also 25 times less severe. The black time series in (b) suggests that an approximation as a stochastic differential equation (SDE) is possible for small $s$. The red trajectories in the figure are in fact from such an SDE. Deriving this equation will be our next step.

We focus on Eq.~(\ref{eq:1dmodel}) and the limit in which the effect of each reset is small. To do this, we write $a(x)=1-sg(x)$, so that the size of the shock is $-sg(x)$, with $s\ll 1$ and where the function $g(x)\geq 0$ does not depend on $s$. Resets are frequent, $r=\lambda/s$, with $\lambda={\cal O}(s^0)$. The reset rate can depend on the state of the process [$\lambda=\lambda(x)$] but we do not always  write this out.

The system can be described by a Kolmogorov equation for the distribution of $x$ at time $t$,
\BE\label{eq:kolmogorov_main}
\frac{\partial}{\partial t} P(x,t)&=&D\frac{\partial^2}{\partial x^2} P(x,t)-\frac{\partial}{\partial x}\left[f(x)P(x,t)\right]\nonumber \\
&&\hspace{-4em} +\int dx' \left[W(x|x') P(x',t)-W(x'|x)P(x,t)\right].
\EE
Here, $W(x|x')$ is the rate of transitioning from $x'$ to $x$ due to resetting. In our setup, we have $W(x|x')=(\lambda/s)\delta\{x-[x'-sg(x')]\}$. For $g(x)=x$ this reduces to the evolution equation for partial resetting in \cite{tal-friedman2022}. Carrying out a Kramers--Moyal expansion in $s\ll 1$ (see End Matters), we find the following Fokker--Planck equation
\BE\label{eq:effective_fpe_main}
\frac{\partial}{\partial t} P(x,t)
&=& D\frac{\partial^2}{\partial x^2} P(x,t) +\frac{1}{2}\lambda s \frac{\partial^2}{\partial x^2} \left[g(x)^2P(x,t)\right]\nonumber \\
&&-\frac{\partial}{\partial x}[f(x)-\lambda g(x)P(x,t)]  +{\cal O}(s^2).
\EE
This describes the It\^o stochastic differential equation 
\be\label{eq:effective_eq}
\dot x = \xi +f(x)-\lambda g(x) +\sqrt{\lambda s}g(x)\eta,
\ee
where $\avg{\eta(t)\eta(t')}=\delta(t-t')$. Equation~(\ref{eq:effective_eq}) is one of the central results of this work, and establishes the diffusion approximation. The first two terms describe the process between resets. The term $-\lambda g(x)$ reflects the {\em mean} change of $x$ due to resets. The noise $\eta$ captures fluctuations of the number of resets per time. 

Equation~(\ref{eq:effective_eq}) can also be derived extending a heuristic for intrinsic noise by Gillespie \cite{GillespieCLE}. Further, the diffusion approximation can be extended to distributed reset sizes. This is described in the End Matters.
\begin{figure}
   \centering
   \vspace{-0.25em}
    \includegraphics[width=1\linewidth]{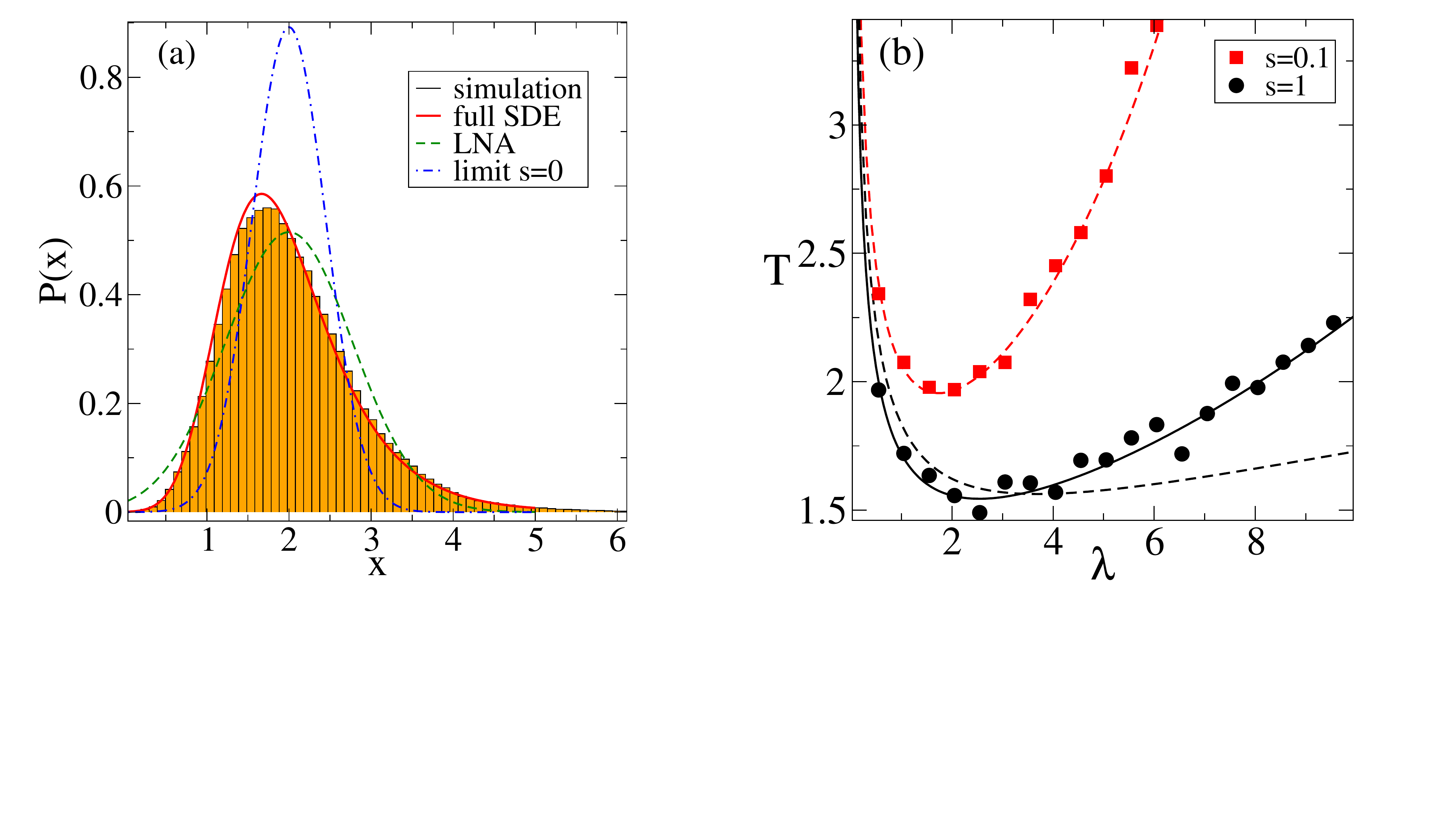}
    \vspace{-5em}
    \caption{(a) Stationary distribution for the model with $f(x)=\alpha+\gamma x$, parameters $s=0.1$, $\alpha=1$, $\lambda=1$, $\gamma=0.5$, $D=0.1$. Orange histogram is from simulations of Eq.~(\ref{eq:1dmodel}) with resetting. Full SDE means Eq.~(\ref{eq:full_main}). LNA refers to the linear-noise appproximation, and `limit $s=0$' to an approximation neglecting the randomness of the resets altogether (for details see the SM). (b) Mean time for a random walk with (partial) resetting to $x_0$ to find a target at the origin ($x_0=1$, $\alpha=\gamma=0$, $D=1$). Results for $s=0.1$ in red, for $s=1$ in black. Markers are from simulations, dashed lines from the diffusion approximation. Solid black line is the exact solution for full resetting from \cite{evans2011diffusion}.
    \label{fig:fig2}}
\end{figure}

{\em Linear example.} As a first test we look at $g(x)=x$ (proportional resets), and $f(x)=\alpha+\gamma x$ ($\alpha$ and $\gamma$ constant). Analytical results exist for diffusion with resetting in potentials, but they take a relatively complicated form \cite{Pal2015}. We therefore proceed with the analysis of the approximate SDE (\ref{eq:effective_eq}). The stationary distribution can be obtained with standard methods \cite{jacobs2010stochastic, gardiner2004handbook}, 
\BE\label{eq:full_main}
P(x)&=& {\cal N}\exp\left[\left(\frac{\gamma-\lambda}{s\lambda}-1\right)\ln(2D+\lambda s x^2)\right]\nonumber \\
&&\times \exp\left[\frac{2\alpha}{\sqrt{2Ds\lambda}}\arctan\left(\sqrt{\frac{s\lambda}{2D}}\,x\right)\right].
\EE
The prefactor ${\cal N}$ ensures normalisation.
We can also carry out a linear-noise approximation \cite{van1992stochastic} replacing the multiplicative noise $g(x)\eta$ in Eq.~(\ref{eq:effective_eq}) with additive noise $g(x^\star)\eta$, where $x^\star$ is the relevant fixed point of $\dot x=f(x)-\lambda g(x)$. Results are shown in Fig.~\ref{fig:fig2} (a). The analytical solution in Eq.~(\ref{eq:full_main}) captures simulation results for $s=0.1$. The linear-noise approximation is still fairly accurate, whereas sending $s\to 0$ (i.e, neglecting all randomness from the resetting) leads to some discrepancies. We show data for $s=0.01$ in the Supplemental Material (SM) \cite{sm} to demonstrate that the diffusion approximation becomes more accurate for smaller $s$.

We can also use the diffusion approximation to study the optimal search problem proposed in \cite{evans2011diffusion}. A random walk is started from $x_0$ and is randomly reset to this position. The goal is to find a target at the origin. Majumdar and Evans showed that there is a reset rate which minimises the expected search time, and they find this rate in closed form. We consider partial resets, $x\to x+s(x_0-x)$, with rate $r=\lambda/s$. The case $s=1$ is the one studied in \cite{evans2011diffusion}. Using the diffusion approximation we obtain an effective stochastic differential equation for $s\ll 1$, and the corresponding backward Fokker--Planck equation. We then find the mean first passage time to reach the origin. Results are shown as dashed lines in Fig.~\ref{fig:fig2}(b). The diffusion approximation captures the mean exit time from simulations for $s=0.1$. For full resetting ($s=1$) we expect and do indeed see deviations. Remarkably, the diffusion approximation continues to capture the existence of an optimal reset rate for $s=1$. 
\medskip

{\em Multi-particle systems.} We now turn to systems with multiple particles, $i=1,\dots,N$. Between resets particles move independently, $\dot x_i = f(x_i)+\xi_i$, with Gaussian white noise variables, $\avg{\xi_i(t)\xi_j(t')}=2D\delta_{ij}\delta(t-t')$. All particles are subject to random (partial) resets $x_i\to a(x_i) x_i$, where $a(x_i)=1-sg(x_i)$. Resets are again triggered with rate $r=\lambda/s$. We assume that $\lambda$ does not depend on the $x_i$. Crucially, in any reset {\em all} particles are reset.

Multi-particle systems with full resets to the origin ($a=1$) were recently studied in \cite{biroli2023extreme}, revealing an interesting `conditional independent and identically distributed' structure (CIID). The distribution of the $x_i$, conditioned on the time of the most recent reset, factorises into copies of the single-particle propagator. This is because a full reset makes the previous history irrelevant, and because the particles move independently between resets. 

Using a system with simultaneous partial resets, we can show that the CIID structure in \cite{biroli2023extreme} is a manifestation of a more general picture. The multi-particle distribution is a product of identical factors if one conditions on the history of reset events. That is to say, if the times at which the partial resets occur are given, then the $\{x_i\}$ become independent and identically distributed. As we will describe, averaging over the common reset history generates correlations between the particles, despite the fact that they move independently between resets.

For $s\ll 1$ we find the diffusion approximation (see End Matters) \be\label{eq:effective_eq_multi}
\dot x_i = \xi_i +f(x_i)-\lambda g(x_i) +\sqrt{\lambda s}g(x_i)\eta.
\ee
The noise variable $\eta$ again encodes the fluctuation of the density of reset events in time.  Crucially, the noise $\eta$ is common to all particles $i$ (it does not carry an index $i$). Equation~(\ref{eq:effective_eq_multi}) shows that, conditioned on a realisation of $\eta$, the $\{x_i\}$ are independent and identically distributed.

To demonstrate that the particles are nonetheless correlated after average over $\eta$, we use the simple case $f(x_i)=\gamma x_i$ and $g(x_i)=x_i$. Writing $\mu\equiv \lambda-\gamma$ (and assuming $\mu>0$), we find for zero initial condition
\BE
x_i(t)&=&\sqrt{2D}\int_0^t \exp\left[\left(-\mu-\frac{1}{2}\lambda s\right)(t-t')\right] \nonumber \\ && \times \exp\left[\sqrt{\lambda s}\left[\nu(t)-\nu(t')\right] \right]\xi_i(t') dt',
\EE
where $\nu(t)=\int^t dt' \eta(t')$. The $\{\xi_i\}$ are independent from one another, so that the $\{x_i\}$ are also independent for any fixed realisation of $\eta$. We can also compute objects such as $\avg{x_i(t)^2|\eta}$ (detailed expressions are given in the SM).  Finally, averaging over $\eta$, we obtain a closed expression for the first non-trivial correlation. As $t\to\infty$ we find
\be
\avg{x_i^2(t) x_j^2(t)}-\avg{x_i(t)^2}^2\to \frac{8D^2\lambda s}{(2\mu-3\lambda s)(2\mu-\lambda s)^2}.
\ee
This quantity is non-zero, confirming the presence of dynamically emerging correlations.
\begin{figure}[t!!!]
\vspace{1em}
   \begin{center}
    \includegraphics[width=1\linewidth]{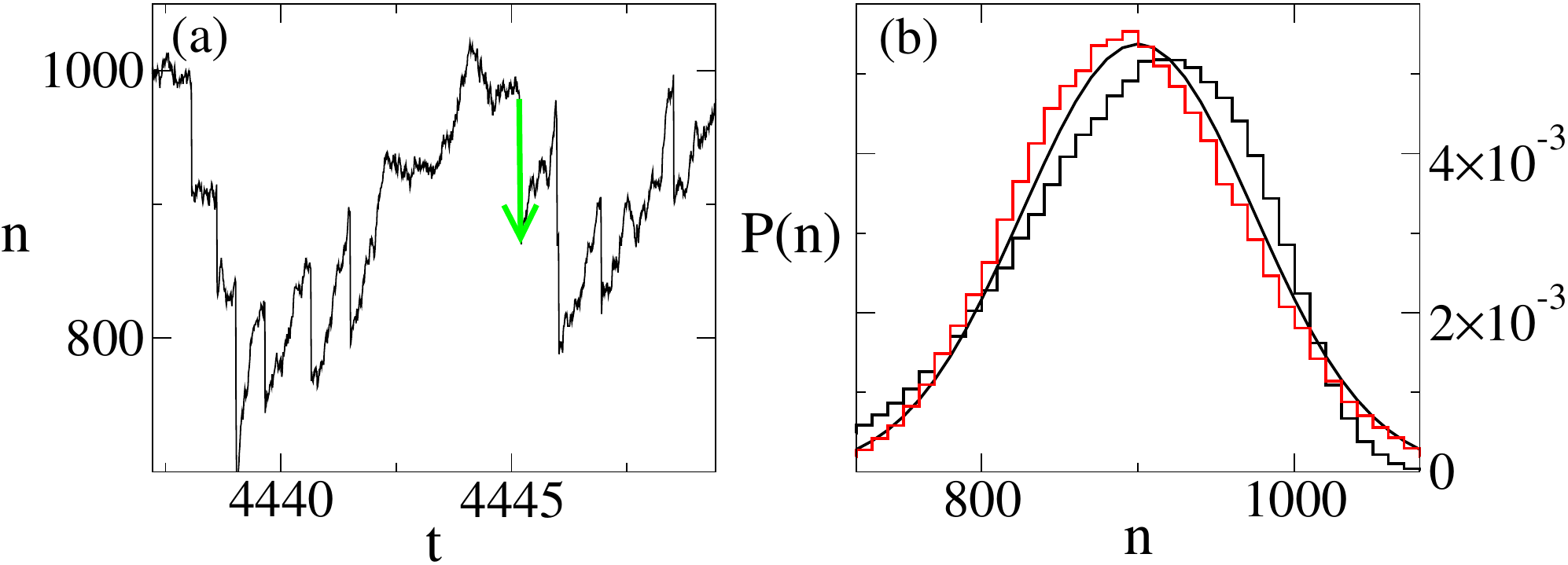}
    \end{center}
    \caption{Individual-based population dynamics [Eq.~(\ref{eq:ibm})] with catastrophes. Panel (a) shows a segment of one realisation ($\lambda=0.1, s=0.1, \Omega=1000$). The green arrow indicates one reset, on average $10\%$ of the popoulation die in such an event. Panel (b) shows the quasi-stationary distribution. The black histogram is from simulations of the individual-based model, the red histogram from simulations of the diffusion approximation [Eq.~(\ref{eq:ibm_sde})]. The smooth black line is from the theory, based on the linear-noise approximation.
    \label{fig:fig3}}
\end{figure}

\medskip

{\em Individual-based model with catastrophes.}
We turn to a simple model of stochastic population dynamics. The population consists of $n(t)$ discrete individuals $A$, with birth and death events of the form
\BE\label{eq:ibm}
A&\longrightarrow& A+A, \nonumber \\
A+A&\longrightarrow& A.
\EE
The first reaction describes proliferation events ($n\to n+1$); the second reaction captures reduction of the population due to competition ($n\to n-1$). We use the rates $T_1(n)=n$ and $T_2(n)=n^2/\Omega$ for these events, where $\Omega$ is a parameter setting the scale of the population size.  In a naive deterministic limit we would have $\dot n = n-n^2/\Omega$, and thus a long-term value of $n=\Omega$. 

The population is now made subject to randomly occurring catastrophes (resets). These happen with rate $r=\lambda/s$, and if a reset occurs each individual in the population independently dies with probability $s$. The combined population and reset dynamics can be simulated using the Gillespie algorithm \cite{Gillespie1976}. An example of a trajectory is shown in Fig.~\ref{fig:fig3}(a). Resets are clearly visible. Between resets the dynamics proceeds via a stochastic sequence of birth and death events. 

There are several sources of randomness in this system. Intrinsic noise comes from the random order and timing of births and deaths. Extrinsic noise results from the random reset times, and from the binomial distribution of catastrophe sizes.

We can write down the master equation for this system. For $\Omega\gg 1$ and $s\ll 1$ we carry out a combined Kramers--Moyal expansion in powers of $1/\Omega$ and an expansion in $s$ (see SM). We then arrive at the following stochastic differential equation for $x=n/\Omega$,

\be\label{eq:ibm_sde}
\dot x = x(1-x)-\lambda x +\sqrt{\frac{x+x^2}{\Omega}}\,\xi+\sqrt{\lambda s x^2+\frac{\lambda}{\Omega} x}\, \eta,
\ee
where $\xi$ and $\eta$ are independent white noise variables, of mean zero and with unit variance, $\avg{\eta(t)\eta(t')}=\avg{\xi(t)\xi(t')}=\delta(t-t')$.
The term $-\lambda x$ in Eq.~(\ref{eq:ibm_sde}) captures the mean effect of catastrophes. The noise $\xi$ describes the randomness of the birth and death dynamics, and $\eta$ comes from the randomness of the resets.

Any realisation of the individual-based model will reach extinction at $n=0$ eventually. The quasi-stationary distribution of $n$ can be estimated analytically from Eq.~(\ref{eq:ibm_sde}) implementing a linear-noise approximation. We compare results for the individual-based model, the diffusion approximation and the linear-noise approximation in Fig.~\ref{fig:fig3}(b), and find reasonable agreement for population size parameter $\Omega=1000$ and for catastrophes in which $10\%$ of individuals are killed on average.
\medskip

{\em Resetting-induced cycles and patterns.}
It is well-known that intrinsic noise can generate quasi-cycles or patterns \cite{mckane_newman, biancalani, butlergoldenfeld}. We now show that there are also resetting-induced cycles and patterns, and that these can be described  using the diffusion approximation.  

We first look at a nonspatial two-species Lotka--Volterra system \cite{mckane_newman},
\BE\label{eq:LV}
\dot x &=&xy-ax, \nonumber \\
\dot y&=&by(1-y/k)-cxy.
\EE
The variables $x(t)$ and $y(t)$ are the abundances of predators and prey, respectively. The parameters $a, b$ and $c$ describe predator decay, raw growth of the prey, and predation effect on the prey, respectively, and $k$ is a carrying capacity for the prey. 

 We subject the dynamics to random catastrophes with rate $r=\lambda/s$, and affecting only the prey population. Upon each reset, the prey population $y$ is reduced by a fraction $s$. In order to highlight the effects of these random catastrophes, we do not include intrinsic noise. In the diffusion approximation, we obtain
\BE\label{eq:LV_diff_approx}
\dot x &=&xy-ax, \nonumber \\
\dot y&=&by(1-y/k)-cxy-\lambda y +\sqrt{\lambda s} y\eta,
\EE
where $\avg{\eta(t)\eta(t')}=\delta(t-t')$. We focus on parameter values for which the deterministic system ($s=0$) converges to a stable fixed point.

\begin{figure}[t!!!]
\vspace{0.25em}
   \begin{center}
    \includegraphics[width=1\linewidth]{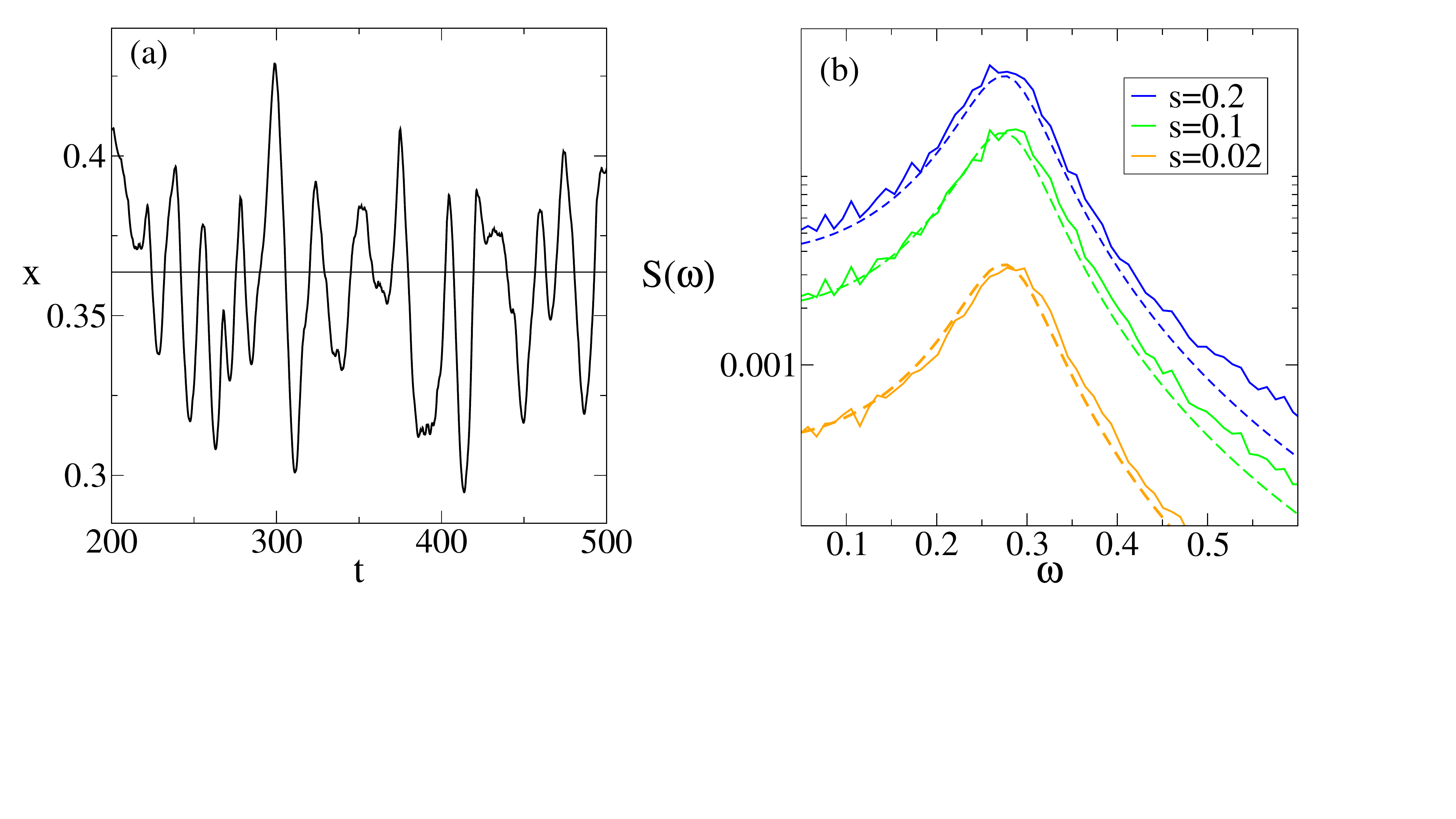}
    \end{center}
    \vspace{-5em}
    \caption{Resetting induced cycles in a predator-prey system [Eq.~(\ref{eq:LV}), $k=1, a=0.1, b=1, c=2.2$]. Panel (a) shows an example of a trajectory (predator time series) for $\lambda=0.1, s=0.1$. Panel (b): Power spectra of the predator time series for different sizes $s$ of the catastrophes. Wiggly lines are from simulations (average over 500 realisations), solid lines from the theory [Eq.~(~\ref{eq:spectrum})].
    \label{fig:fig4}}
\end{figure}

An example of time series $x(t)$ from the system in Eq.~(\ref{eq:LV}) with random resets is shown in Fig.~\ref{fig:fig4}(a). The horizontal line is the deterministic fixed point of Eq.~(\ref{eq:LV_diff_approx}) by setting $s=0$. Resetting-induced oscillations are clearly visible in the figure. 

To study these cycles we follow \cite{mckane_newman}. We identify the fixed point $(x^\star, y^\star)$ of Eqs.~(\ref{eq:LV_diff_approx}) for $s=0$, and linearise about this fixed point for small but non-zero writing $x=x^\star+u$ and $y=y^\star+v$. Making the linear-noise approximation we find
\BE
\dot u &=& J_{11} u + J_{12} v, \nonumber \\
\dot v &=& J_{21} u + J_{22} v+\sqrt{\lambda s}y^\star\eta,
\EE
where the $J_{ij}$ are the entries of the Jacobian at the fixed point. We then carry out a Fourier transform, and obtain the power spectra of the fluctuations about the fixed point. For the predator species we find
\be\label{eq:spectrum}
S(\omega)=\avg{|\tilde u(\omega)|^2}=\frac{\lambda s (y^\star)^2 J_{12}^2}{(\omega^2-\Omega_0^2)^2+\Gamma^2\omega^2},
\ee
where $\Omega_0^2=J_{11}J_{22}-J_{12}J_{21}$ and $\Gamma^2=(J_{11}+J_{22})^2$. As seen in Fig.~\ref{fig:fig4}(b) this agrees with simulation of the original system [Eqs.~(\ref{eq:LV}), plus resets]. Eq.~(\ref{eq:spectrum}) shows that the amplitude of the oscillations scales at $\sqrt{\lambda s}$.  This confirms again that the cycles are resetting-induced.

To demonstrate catastrophe-induced patterns we use the Levin--Segel model of plankton-herbivore dynamics \cite{levin1976hypothesis}. The model describes spatially varying fields, $\psi(\bx,t)$ and $\phi(\bx,t)$ for the local densities of plankton and herbivore respectively. In our setup space is discrete, modelling patches arranged on a spatial grid. We use the notation in \cite{butler2009robust, butlergoldenfeld}, and  we allow for catastrophes with rate $\lambda=r/s$, occurring as a random process independently at patch, and each removing a fraction $s$ of the local plankton population. The diffusion approximation is then
\BE\label{eq:diff_approx_space}
\frac{\partial}{\partial t} \psi&=&D_\psi \Delta \psi + \frac{1}{2}\psi+\frac{1}{2}\psi^2-p\psi\phi-\lambda \psi +\sqrt{\lambda s}\psi\eta, \nonumber \\
\frac{\partial}{\partial t}\phi&=&D_\phi \Delta\phi+\phi\psi-\frac{1}{2}\phi^2,
\EE
with $\Delta$ the lattice Laplacian. The terms not containing $\lambda$ are as in \cite{butler2009robust, butlergoldenfeld} and describe the model without resets. The term $-\lambda\psi$ captures the mean effect of resets. The noise $\eta$ comes from the randomness due to resets. 

\begin{figure}[t!!!]
   \begin{center}   
     \includegraphics[width=0.85\linewidth]{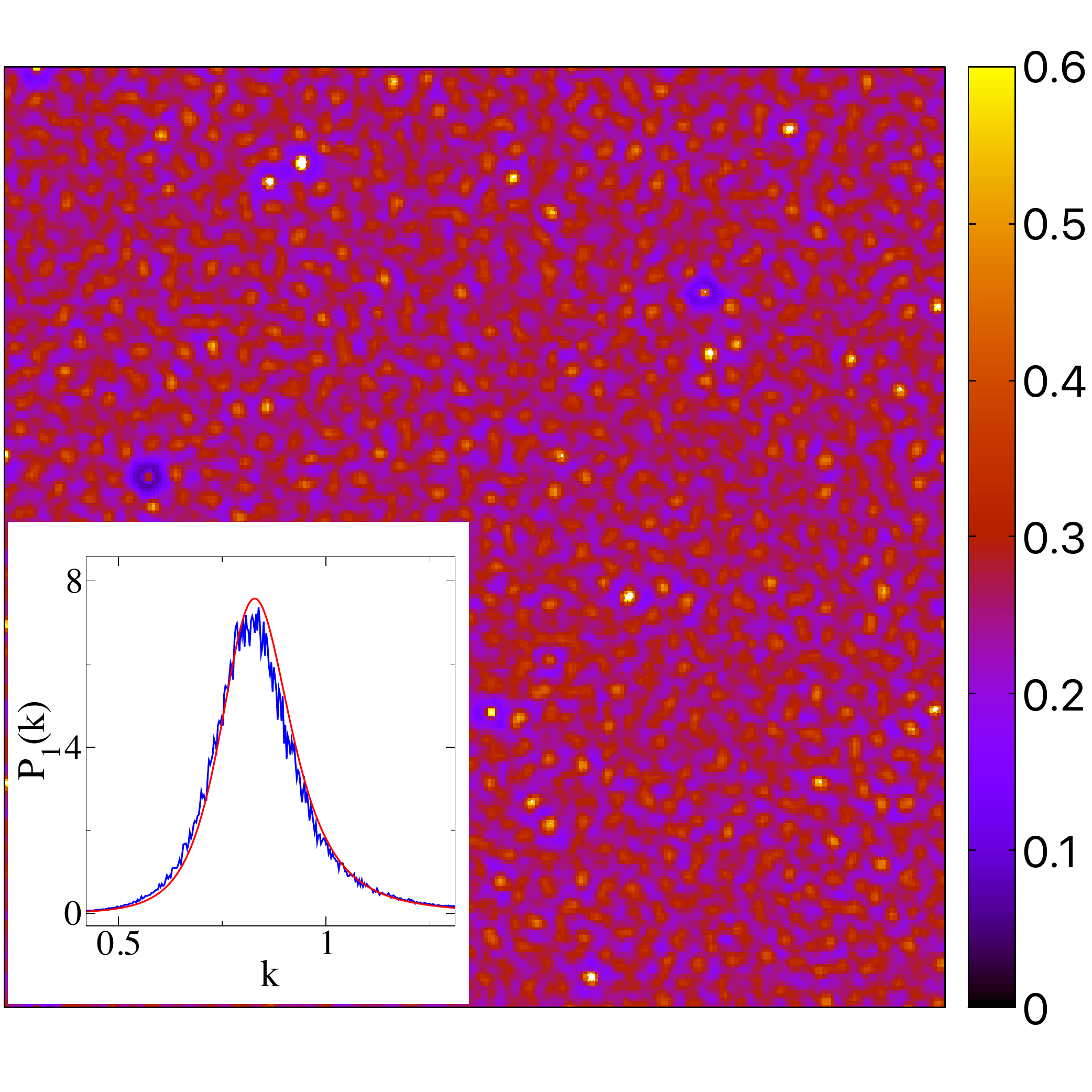}
    \end{center}
    \vspace{-3em}
    \caption{Main panel: Resetting induced pattern (plankton species $\psi$) in the Levin--Segel model in two dimensions. Inset: Spectrum of flucutations $\avg{|\delta \tilde \psi(k,\omega=0)|^2}$ for the model in one dimension as a function of the wavenumber $k$. Wiggly  line from simulations (average over 300 realisations), solid red line from the diffusion and linear-noise approximations. Parameters: $b=d=e=1/2$, $p=1$, $D_\psi=0.1$, $D_\phi=2.5$, $\lambda=0.1$, $s=0.05$. Model operates on discrete patches ($\Delta x=1$), lateral extent in the main panel is $256\times 256$.
    \label{fig:fig5}}
\end{figure}

We focus on parameter values such that Eqs.~(\ref{eq:diff_approx_space}) approach a spatially homogeneous fixed point for $s=0$. In Fig.~\ref{fig:fig5} we show an example of a resetting-induced pattern in two dimensions.

A formal analysis can be carried out starting from Eqs.~(\ref{eq:diff_approx_space}). After carrying out a linear-noise approximation and an inversion in Fourier space of finds the spectra of fluctuations $\delta\psi$ and $\delta\phi$ about the deterministic homogeneous fixed point \cite{sm}. We show an example of a spectrum of plankton density fluctuations for the model in one spatial dimension in the inset of Fig.~\ref{fig:fig5}, focusing on frequency $\omega=0$. The peak at a nonzero wavenumber confirms the presence of resetting-induced patterns. 

{\em Conclusions.}
We have developed a diffusion approximation for extrinsic noise in systems with frequent small random resets or catastrophes. Using the approximation we have computed stationary distributions and escape times, and we have shown how dynamically induced correlations are captured. We have also demonstrated that resetting can induce quasi-cycles and quasi-patterns.

Bearing in mind that the diffusion approximation is routinely used across fields including chemical reaction systems \cite{GillespieCLE}, population genetics \cite{ewens2004mathematical}, and operations research \cite{glynn}, we believe the approximation can be a useful tool for systems with resetting. In particular the approximation allows analytical progress for systems which are not exactly solvable, or where closed-form solutions are intricate. For example, it would be interesting to study multi-particle systems with batch resetting \cite{demauro}.

To close we note that first experimental tests of the theory of resetting have only been conducted very recently \cite{tal2020experimental, besga, faisant, biroli2025experimental}. It is conceivable that predictions from the diffusion approximations can be tested in future experiments.
\medskip

\acknowledgements
	I acknowledge partial financial support from the Agencia Estatal de Investigación and Fondo Europeo de Desarrollo Regional (FEDER, UE) under projects APASOS (PID2021-122256NB-C22) and COSASTI (PID2024-157493NB-C22), and the Mar\'ia de Maeztu programme for Units of Excellence, CEX2021-001164-M, funded by MCIN/AEI/10.13039/501100011033.

    The data that support the findings of this article are available \cite{data}.

\clearpage
\newpage

\onecolumngrid
\begin{center}
    {\bf End Matters} 
\end{center}
\begin{appendix}
In these End Matters we give the detailed derivation of the diffusion approximation for processes with a single variable (Appendix \ref{app:single}) and with multiple variables (Appendix \ref{app:multi}).

\section{Diffusion approximation for single-variable processes}\label{app:single}
\subsection{Setup}
Consider a single real-valued variable $x(t)$ undergoing the process $\dot x(t)= f[(x(t)]+\xi(t)$, where $\xi(t)$ is zero-mean Gaussian white noise of the form $\avg{\xi(t)\xi(t')}=2D\delta(t-t')$,
with $D\geq 0$ a constant. Additionally, with rate $r$ the particle is partially reset as $x\to x-sg(x)$, where $g(\cdot)$ is some function. We write out the theory for $g(x)\geq 0$. The rate $r$ can depend on position, but we do not always spell this out explicitly.

\subsection{Kramers--Moyal expansion}\label{sec:KME}
The distribution $P(x,t)$ of $x$ at time $t$ fulfills the evolution equation
\be\label{eq:kolmogorov}
\frac{\partial}{\partial t} P(x,t)=D\frac{\partial^2}{\partial x^2} P(x,t)-\frac{\partial}{\partial x}\left[f(x)P(x,t)\right]+\int dx' \left[W(x|x') P(x',t)-W(x'|x)P(x,t)\right],
\ee
where $W(x|x')$ is the rate of transitioning to $x$ due to resetting if the particle is currently at position $x'$. In our setup we have $W(x|x')=\frac{\lambda}{s}\delta\{x-[x'-sg(x')]\}$.
\medskip

Carrying out a Kramers--Moyal expansion of the integral term in Eq.~(\ref{eq:kolmogorov}), we have
\BE\label{eq:KME}
\frac{\partial}{\partial t} P(x,t)&=&D\frac{\partial^2}{\partial x^2} P(x,t)-\frac{\partial}{\partial x}\left[f(x)P(x,t)\right]+\sum_{n=1}^\infty\left(-\frac{\partial }{\partial x}\right)^n \left[D_n(x) P(x,t)\right],
\EE
with the jump moments $D_n(x)=\int dx'\, W(x'|x) (x'-x)^n$ for $n=1,2,\dots$. In our model $D_1(x)=-\lambda g(x)$ and $D_2(x)=\lambda s g(x)^2$. 
\medskip

Truncating the series in Eq.~(\ref{eq:KME}) to include only Liouville and diffusion terms we have
\BE\label{eq:KME}
\frac{\partial}{\partial t} P(x,t)&=&D\frac{\partial^2}{\partial x^2} P(x,t)-\frac{\partial}{\partial x}\left[f(x)P(x,t)\right] +\frac{1}{2}\lambda s \frac{\partial^2}{\partial x^2} \left[g(x)^2P(x,t)\right]-\frac{\partial}{\partial x}\left\{-\lambda g(x)P(x,t)\right\}. 
\EE
This describes a process of the form
\be\label{eq:effective_eq_sm2}
\dot x(t) = \xi(t) +f[x(t)]-\lambda g[x(t)] +\sqrt{\lambda s}\, g[x(t)]\,\eta(t),
\ee
with $\avg{\eta(t)\eta(t')}=\delta(t-t')$.
\subsection{Heuristic derivation}\label{sec:heuristic}
Equation~(\ref{eq:effective_eq_sm2}) can also be derived extending an argument originally used for intrinsic noise by Gillespie \cite{GillespieCLE}. 

We discretise time, and look at a short time interval $\Delta t$. Gillespie's approach assumes that $x$ stays constant during this time (changes are implemented at the end of the interval). The number of resets $m$ during the interval is Poisson-distributed with parameter $\Delta t\lambda/s$. Thus, $m$ has mean and variance both equal to $\Delta t\lambda/s$. 

For $s\ll 1$ (i.e., frequent resets), the number of resets in the interval will be much larger than one ($m\gg 1$). Hence, we can approximate $m$ with a Gaussian random variable. This Gaussian is taken to have the same mean and variance as the above Poissonian, i.e., it has mean and variance both equal to $\Delta t\lambda/s$. 

The net change of $x$ is $-sg(x)$ in each reset. Therefore, the mean change of $x$ in the interval $\Delta t$ due to resets is $-sg(x)\avg{m}=-sg(x)\Delta t\lambda/s=-\lambda g(x) \Delta t$, and (to linear order in $\Delta t$) the variance of the reduction is $[sg(x)]^2\mbox{Var}(m)=[sg(x)]^2 \Delta t\lambda/s=\lambda s g(x)^2 \Delta t$. Making the Gaussian approximation, and restoring the continuous-time limit by sending $\Delta t\to 0$, this leads to Eq.~(\ref{eq:effective_eq_sm2}).

\subsection{Reset with random amplitude}\label{sec:random_amplitude}
Now consider a generalisation, in which reset occur with rate $r=\lambda/s$ as before, but where the amplitude of the resets fluctuates. For simplicity we focus on proportional resets, i.e., $g(x)=x$. We assume that a reset occurs as $x\to \left[1-s(1+z)\right]x$, where $z$ is a random variable of mean zero and variance $\sigma^2$. The value of $z$ is drawn independently each time a reset is triggered.

We extend the heuristic in Sec.~\ref{sec:heuristic}. In a given time interval of length $\Delta t\ll 1$, the variable $x$ the changes by $\Delta x = - xms -sx\sum_{\ell=1}^m z_\ell$,
where $m$ is again a Poissonian random variable with mean and variance $\lambda \Delta t/s$ and where the $z_\ell$ are {\em iid} with mean zero and variance $\sigma^2$. We have (similar to before), $\avg{\Delta x}=-\lambda x \Delta t$. Using the law of total variance, we also find
\be
\mbox{Var}(\Delta x)=\mbox{E}_m [\mbox{Var}(\Delta x|m)]+\mbox{Var}_m [\mbox{E}(\Delta x|m)] 
=s^2\sigma^2x^2\mbox{E}_m[m]+x^2s^2\mbox{Var}_m[m] 
=sx^2\lambda\Delta t (1+\sigma^2).
\ee
Here $E_m[\dots]$ stands for an average over $m$, and $\mbox{Var}_m(\cdots)$ for the corresponding variance.
\medskip

This means that we have a process of the following form [with $\avg{\eta(t)\eta(t')}=\delta(t-t')$],
\be\label{eq:effective_eq2}
\dot x(t) = \xi(t) +f[x(t)]-\lambda x(t) +\sqrt{\lambda s (1+\sigma^2)} x(t)\eta(t).
\ee

\section{Multi-particle system}\label{app:multi}
\subsection{Setup}
Now consider multiple walkers with positions $x_i$, $i=1,\dots,N$. Between resets the particles move independently according to $\dot x_i(t)= f[x_i(t)]+\xi_i(t)$,
where the $\xi_i(t)$ are Gaussian white noise (mean zero) with $\avg{\xi_i(t)\xi_j(t')}=2D\delta_{ij}\delta(t-t')$.
In particular $\xi_i$ and $\xi_j$ are independent from one another for $i\neq j$. Resets occur as before with rate $\lambda/s$. All particles are (partially) reset simultaneously when a reset is triggered, i.e., $
x_i \to x_i-sg(x_i)$ for all $i$.

\subsection{Derivation of the diffusion approximation}
\subsubsection{Heuristic derivation}
We again discretise time into small intervals, and we assume that the $x_i$ are updated only at the end of the interval. The number of resets in an interval is again a Poisson random variable $m$ with mean and variance $\Delta t\lambda/s$. Importantly, all particles experience the exact same number of resets in the interval. As before, we approximate $m$ with a Gaussian random variable for $s\ll 1$, with mean and variance both equal to $\Delta t\lambda(x)/s$. 

The net change of $x_i$ is $-sg(x_i)$ in each reset. Therefore, the expected change of $x_i$ in the interval $\Delta t$ due to resets is $-\lambda g(x_i) \Delta t$. To linear order in $\Delta t$ the variance of the reduction of $x_i$ is $[sg(x_i)]^2\mbox{Var}(m)=\lambda s g(x_i)^2 \Delta t$. Importantly, all particles $x_i$ experience the same fluctuations, i.e., if the number of resets is lower/higher than average for one particle than it is lower/higher than average for all particles.  Restoring the continuous-time limit by sending $\Delta t\to 0$ we can describe the effect of resetting on particle $i$ as $-\lambda g(x_i)+\sqrt{\lambda s} g(x_i)\eta$, where the noise term $\eta$ is common to all $i$. This leads to the diffusion approximation $
\dot x_i(t) = f[x_i(t)]-\lambda x_i(t) +\xi_i(t)+\sqrt{\lambda s}\, g(x_i) \eta(t)$.
\subsubsection{Kramers--Moyal expansion}
 Writing $\bx=(x_1,\dots,x_N)$, we have the transition rates
$
W(\bx|\bx')=\frac{\lambda}{s}\prod_{i=1}^N\delta\{x_i-[x_i'-sg(x_i')]\}$
for the reset events. Carrying out a Kramers--Moyal expansion and keeping only terms up to Fokker--Planck order then results in
\BE\label{eq:KME_multi}
\frac{\partial}{\partial t} P(\bx,t)&=&D\sum_i\frac{\partial^2}{\partial x_i^2} P(\bx,t)-\sum_i\frac{\partial}{\partial x_i}\left[f(x_i)P(\bx,t)\right] \nonumber \\
&&-\sum_i\frac{\partial}{\partial x_i}\left\{[f(x_i)-\lambda g(x_i)]P(\bx,t)\right\} +\frac{1}{2}\lambda s\sum_{ij} \frac{\partial^2}{\partial x_i\partial x_j} \left[g(x_i)g(x_j)P(\bx,t)\right].
\EE
The corresponding set of stochastic differential equations is
$\dot x_i(t) = f[x_i(t)]-\lambda x_i(t) +\xi_i(t)+\nu_i(t)$,
with $
\avg{\nu_i(t)\nu_j(t)}=\lambda s g[x_i(t)]g[x_j(t)]\delta(t-t')$.
We stress that there is no factor $\delta_{ij}$ on the right. This structure means that the $\{\nu_i(t)\}$ can be written in the form $\nu_i(t)=\sqrt{\lambda s} g[x_i(t)]\eta(t)$, where the Gaussian standard noise $\eta(t)$ is common to all $i$. 

\end{appendix}
\end{document}

\title{A diffusion approximation for systems with frequent weak resetting\\~\\  --- Supplemental Material ---}

\author{Tobias Galla}
\email{tobias.galla@ifisc.uib-csic.es}
\affiliation{Instituto de F{\' i}sica Interdisciplinar y Sistemas Complejos IFISC (CSIC-UIB), 07122 Palma de Mallorca, Spain}

\maketitle
 \tableofcontents

\section{Example: Linear single-variable process}
In this section we provide further details and additional result for the model with a single degree of freedom $x(t)$ undergoing the process $\dot x(t)= f[(x(t)]+\xi(t)$, where $\xi(t)$ is zero-mean Gaussian white noise of the form $\avg{\xi(t)\xi(t')}=2D\delta(t-t')$,
with $D\geq 0$ a constant. 

The particle is partially reset with rate $r$, and resets take the form $x\to x-sg(x)$, where $g(\cdot)$ is a fixed function. In what follows we assume $g(x)\geq 0$, but the equations can  be adapted. The rate $r$ can depend on position, but we do not always write this down explicitly.

The Fokker--Planck equation in the diffusion approximation is derived in the End Matters in the main paper [Eq.~(A3)], we repeat it here for completeness,
\BE\label{eq:FPE_sm}
\frac{\partial}{\partial t} P(x,t)&=&D\frac{\partial^2}{\partial x^2} P(x,t)-\frac{\partial}{\partial x}\left[f(x)P(x,t)\right] +\frac{1}{2}\lambda s \frac{\partial^2}{\partial x^2} \left[g(x)^2P(x,t)\right]-\frac{\partial}{\partial x}\left\{-\lambda g(x)P(x,t)\right\}. 
\EE
This describes a process of the form
\be\label{eq:sde_sm}
\dot x(t) = \xi(t) +f[x(t)]-\lambda g[x(t)] +\sqrt{\lambda s}\, g[x(t)]\,\eta(t),
\ee
with $\avg{\eta(t)\eta(t')}=\delta(t-t')$.

\subsection{Stationary solution of the effective Fokker-Planck equation}
The stationary solution of the Fokker--Planck equation (\ref{eq:FPE_sm}) fulfills
\be
-\partial_x [A(x) P(x)]+\frac{1}{2}\partial_x^2 [B(x)P(x)]=0,
\ee
where
\BE
A(x)=f(x)-\lambda g(x), \nonumber \\
B(x)=2D+\lambda s g(x)^2.
\EE
Assuming reflecting boundary conditions, we then have the following stationary distribution (see e.g \cite{gardiner2004handbook, jacobs2010stochastic}),
\BE\label{eq:stat_distrib_sm}
P(x)&=&{\cal N}_1\frac{1}{B(x)}\exp\left(2\int^x dy \frac{A(y)}{B(y)}\right),
\EE
with ${\cal N}_1$ a suitable normalisation constant. 
\medskip

\noindent\underline{Example: $f(x)=\alpha+\gamma x$}\\
The remaining integral in Eq.~(\ref{eq:stat_distrib_sm}) cannot be evaluated in closed form for arbitrary functions $f(x)$ and $g(x)$. We therefore restrict ourselves to affine linear functions
\be
f(x)=\alpha +\gamma x,
\ee
where $\alpha$ and $\gamma$ are constants. We also set $g(x)=x$ (proportional resetting), and assume that $\lambda>\gamma$.

We then have
\BE
\int^x dy \frac{\alpha+(\gamma-\lambda) y}{2D+\lambda s y^2}=\frac{\alpha}{\sqrt{2Ds\lambda}}\arctan\left(\sqrt{\frac{s\lambda}{2D}}\,y\right)
+\frac{\gamma-\lambda}{2s\lambda}\ln\left[2D+s\lambda y^2\right],
\EE
and hence
\BE\label{eq:full_sm}
P(x)&=& {\cal N}\exp\left[\left(\frac{\gamma-\lambda}{s\lambda}-1\right)\ln(2D+\lambda s x^2)\right] \times \exp\left[\frac{2\alpha}{\sqrt{2Ds\lambda}}\arctan\left(\sqrt{\frac{s\lambda}{2D}}\,x\right)\right].
\EE
This is the expression used to generate the theory line in Fig.~2 in the main text.
\medskip

\noindent\underline{Limit $s\to 0$}\\
For $s\ll 1$ we have
\be
\ln(2D+\lambda s x^2)=\ln(2D)+\ln\left(1+\frac{\lambda sx^2}{2D}\right)\approx \ln(2D)+\frac{\lambda sx^2}{2D}.
\ee
Similarly,
\be
 \frac{2\alpha}{\sqrt{2Ds\lambda}}\arctan\left(\sqrt{\frac{s\lambda}{2D}}\,x\right)\approx \frac{2\alpha}{\sqrt{2Ds\lambda}}\left(\sqrt{\frac{s\lambda}{2D}}\,x\right)=\frac{\alpha}{D}x.
\ee
For $s\to 0$ we therefore find
\be\label{eq:P_s_eq_0}
P_{s\to 0}(x)={\cal N}_2 \exp\left[-\frac{\lambda-\gamma}{2D}\left(x-\frac{\alpha}{\lambda-\gamma}\right)^2\right],
\ee
with ${\cal N}_2$ a normalisation constant. 

This distribution is well defined only for $\lambda>\gamma$, and is then a Gaussian centred on $x=\alpha/(\lambda-\gamma)$, and with variance $D/(\lambda-\gamma)$. 
This the stationary distribution in the limit $s\to 0$, i..e, infinitely fast infinitely small resets. The dynamics in Eq.~(4) in the main paper [with $f(x)=\alpha+\gamma x$] is then $\dot x = \alpha -(\lambda-\gamma)x+\xi$. Setting $y=x-\alpha/(\lambda-\gamma)$, this can be re-written as a standard Ornstein-Uhlenbeck process $\dot y = -(\lambda-\gamma)y+\xi$, and Eq.~(\ref{eq:P_s_eq_0}) is its well-known stationary distribution.

\medskip

\noindent \underline{Linear noise approximation}\\
The linear-noise approximation (LNA) \cite{van1992stochastic} consists of replacing $x(t)$ in the factor multiplying $\eta$ in the SDE (\ref{eq:sde_sm}) with the fixed-point value $x^*=\alpha/(\lambda-\gamma)$. We then have
\be\label{eq:LNA}
\dot x = \alpha+(\gamma-\lambda) x +\sqrt{2D+\lambda s \left(\frac{\alpha}{\lambda-\gamma}\right)^2}\zeta,
\ee
with standard Gaussian white noise $\zeta$. This is again an Ornstein-Uhlenbeck process. Its stationary distribution is
\be\label{eq:P_LNA}
P_{\rm LNA}(x)={\cal N}_3 \exp\left[-\frac{\lambda-\gamma}{2D+\lambda s \left(\frac{\alpha}{\lambda-\gamma}\right)^2}\left(x-\frac{\alpha}{\lambda-\gamma}\right)^2\right].
\ee

\begin{figure}
    \centering
    \includegraphics[width=0.5\linewidth]{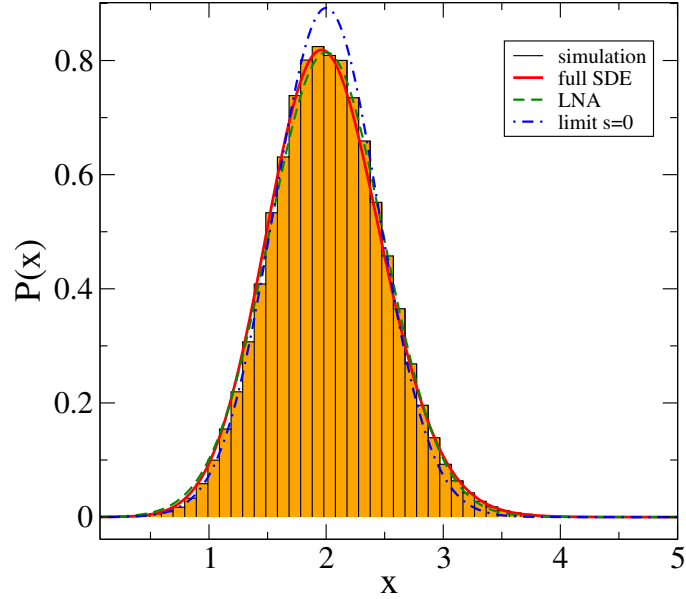}
    \caption{Stationary distribution for the model with $f(x)=\alpha+\gamma x$. Parameters: $\alpha=1$, $\lambda=1$, $\gamma=0.5$, $D=0.1$, $s=0.01$. Simulations run up to time $20,000$, time step $0.001$. Data sampled from time $1000$ onwards. Full SDE means Eq.~(\ref{eq:full_sm}). LNA refers to Eq.~(\ref{eq:P_LNA}) and `limit $s=0$' to Eq.~(\ref{eq:P_s_eq_0}).
    \label{fig:stat_distrib_sm}}
\end{figure}

We compare the validity of the approximations in Eqs.~(\ref{eq:full_sm}) and (\ref{eq:P_LNA}) against simulations in Fig.~2 of the main paper (for $s=0.1$). Results for $s=0.01$ are shown in Fig.~\ref{fig:stat_distrib_sm} in this Supplement.

\subsection{Resets with distributed amplitude}

Further, we have also tested the theory for resets with random amplitude (Appendix A.3 in End Matters).  Resets are now of the form $x\to \left[1-s(1+z)\right]x$, where $z$ is a random variable of mean zero and variance $\sigma^2$. The value of $z$ is drawn independently each time a reset is triggered.

\begin{figure}
    \centering
  \includegraphics[width=0.4\linewidth]{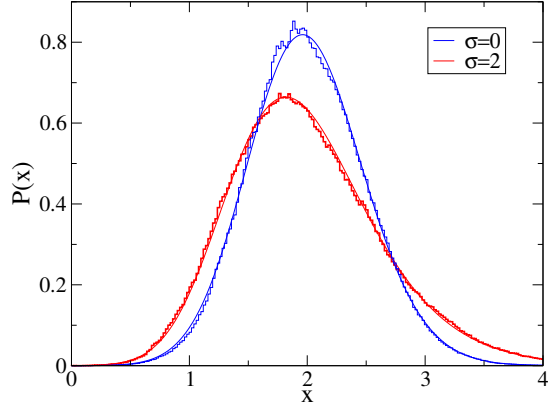}
    \caption{Stationary distribution for the model with $f(x)=\alpha+\gamma x$ and random resetting amplitude. Parameters: $\alpha=1$, $\lambda=1$, $\gamma=0.5$, $D=0.1$, $s=0.01$. Data is shown for $\sigma=0$ and $\sigma=2$. Wiggly lines are from simulations run up to time $20,000$, time step $0.0001$. Data sampled from time $1000$ onwards. The smooth lines are from Eq.~(\ref{eq:full_sm}), with $s$ replaced by $s(1+\sigma^2)$ to account for randomly distributed reset amplitudes (see also Appendix A.3 in End Matters).
    \label{fig:random_amount}}
\end{figure}

Results are shown in Fig.~\ref{fig:random_amount}, where $z$ is a uniform random variable with mean zero and variance $\sigma^2$. As seen from the data, the theory adequately describes the stationary distribution. We show results for $\sigma=0$ (no randomness in the jump sizes) and $\sigma=2$, to show that the theory captures the effects of randomly distributed reset amplitudes. As expected, the stationary distribution is broader for random reset amplitudes than for fixed proportional reset sizes.

\
\subsection{First-passage time}
We compute the first passage time for the process to leave the interval $[-1,1]$ starting from an initial position $x=x_0$.

For $s\ll 1$ [and setting $g(x)=x$] we have from Eq.~(\ref{eq:FPE_sm})
\BE\label{eq:effective_fpe_2}
\frac{\partial}{\partial t} P(x,t)
&=& \frac{1}{2}\frac{\partial^2}{\partial x^2}\left[\left(2D+\lambda s x^2\right) P(x,t)\right] -\frac{\partial}{\partial x} \left\{[f(x)-\lambda x]P(x,t)\right\}.  
\EE
We use again the definitions
\BE
A(x)=f(x)-\lambda x, \nonumber \\
B(x)=2D+\lambda s x^2.
\EE
The mean first passage time $T(y)$ starting from a point $y$ then fulfills \cite{jacobs2010stochastic} (with a sign error corrected relative to this source)
\be
\frac{1}{2}B(y)T''(y)+A(y)T'(y)+1=0.
\ee
We focus again on $f(x)=\alpha+\gamma x$. Then we have
\be\label{eq:T_backward}
\frac{1}{2}[2D+\lambda s y^2]T''(y)+[\alpha+(\gamma-\lambda)y]T'(y)+1=0,
\ee
which is to be solved subject to the boundary conditions $T(-1)=T(1)=0$. In principle, a closed form solution can be obtained using expressions given for example in \cite{jacobs2010stochastic}, but these involve nested integrals and are not very informative. We therefore solve the boundary value problem numerically using Mathematica. 
\begin{figure}[t!!!]
\vspace{1em}
   \begin{center}
    \includegraphics[width=0.5\linewidth]{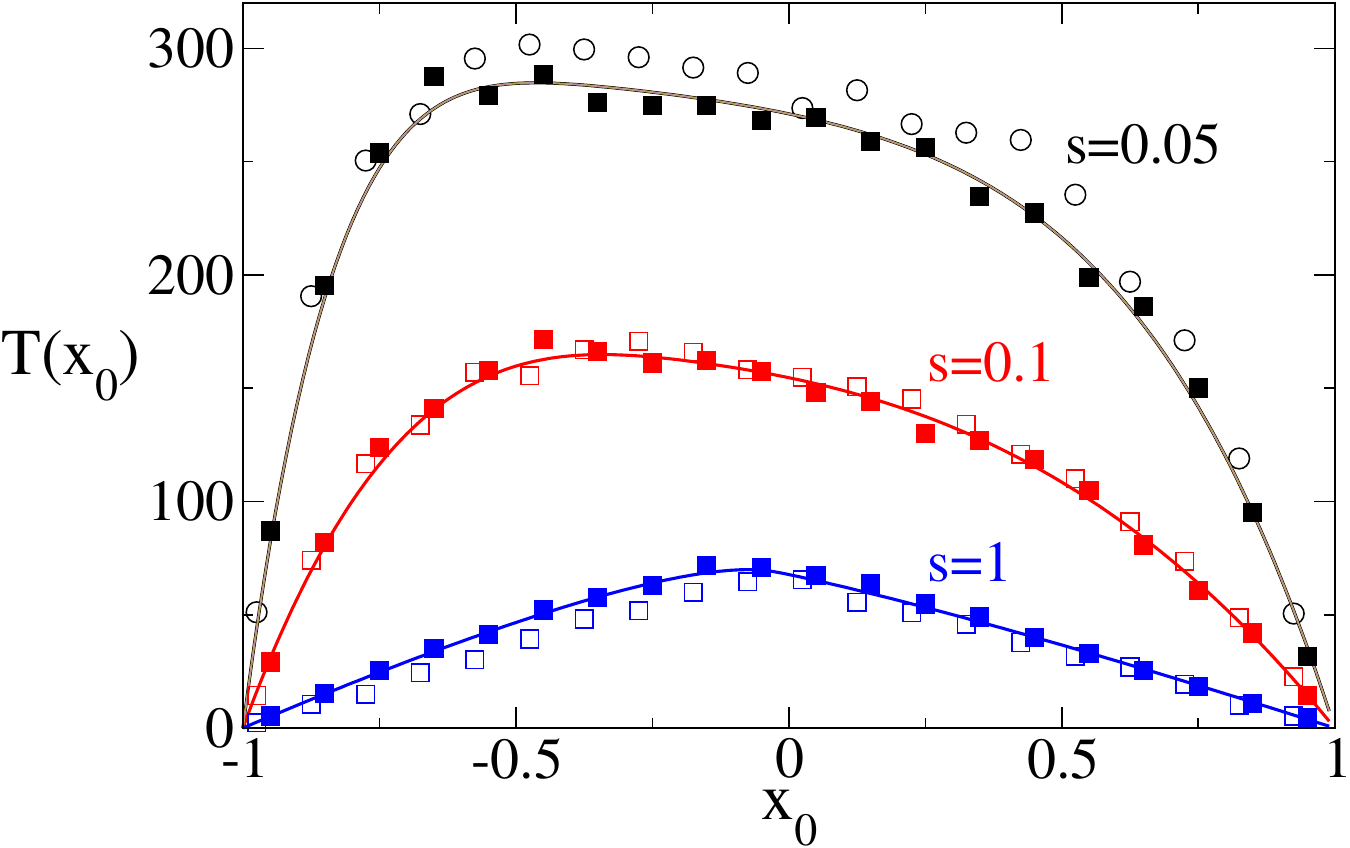}
    \end{center}
    \caption{Mean exit time from the interval $[-1,1]$ as a function fo the starting point $x_0$. Parameters: $\alpha=0.02, \lambda=0.55, \gamma=0.5$. We set $D=0$ in panel so that all stochastic effects come from the resetting. Solid lines are a numerical solution of Eq.~(\ref{eq:T_backward}), with boundary conditions $T(1)=T(-1)=0$. Full symbols are from numerical integration of the SDE (\ref{eq:sde_sm}) with $f(x)=\alpha+\gamma x$, open symbols from simulations of the original process. Each data point is from 1000 independent realisations, with time step $0.0001$.
    \label{fig:exit_time}}
\end{figure}
\medskip

Results are shown in Fig.~\ref{fig:exit_time}. We include simulation results for the stochastic differential equation [Eq.~(\ref{eq:sde_sm})] to test the validity of the numerical solution of the backward Fokker--Planck equation. Predictions from the backward Fokker--Planck equation agree well with simulations. We attribute deviations at $s=1$ to inaccuracies of the Gaussian approximation. 

We also find small deviations between theory and simulations of the full model at $s=0.05$ and at intermediate starting points (when escape times are relatively long). We attribute this to discretisation errors and to inaccuracies of the diffusion approximation, noting that these aggregate over time and thus become more pronounced as escape times become longer.

\begin{figure}
\vspace{5em}
    \centering
    \includegraphics[width=0.5\linewidth]{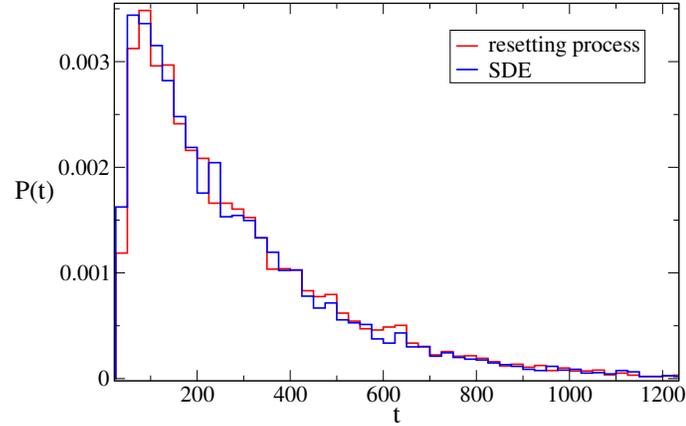}
        \caption{Distribution of escape times time from the interval $[-1,1]$ if the process is started at $x=x_0=-0.25$. Model parameters are $s=0.05$, $\alpha=0.02, \lambda=0.55, \gamma=0.5$, and we have set $D=0$. Data is from $10000$ realisations, integration time step $0.001$.   \label{fig:hist_T_esc}}
\end{figure}

To test this further we plot the distribution of escape times for a starting point $x_0=-0.25$ in Fig.~\ref{fig:hist_T_esc}. We show data from the integration of the original process, and from integrating the effective stochastic differential equation (\ref{eq:sde_sm}). As seen from the figure, the stochastic differential equation reproduces the distribution of escape times to a reasonable accuracy. Mean escape times from these histograms are approximately $291.8$ for the original model, and approximately $282.4$ for the diffusion approximation (so approximately a $3\%$ deviation, some of which will be statistical). The value predicted from a numerical solution of the backward Fokker--Planck equation is $280.7$.

\subsection{Optimal search}
We consider a problem from \cite{evans2011diffusion}. The authors studied a simple random walk starting from position $x_0$, subject to random resetting to $x_0$ at rate $r$. The authors then calculated the mean time to arrive at the origin (mean first passage time) for starting position is $x_0$, and find
\be
T(x_0)=\frac{1}{r}\left(e^{\sqrt{r/D}x_0}-1\right).
\ee
The black line in Fig.~2(b) of the main paper is obtained from this equation (with $r=\lambda/s$). For fixed $D$ and $x_0\neq 0$, this function has a mimimum at intermediate $r$, thus indicating that there is an optimal resetting rate that minimises the time to find a target at the origin.

\medskip

We now consider an analogous problem for a random walk with frequent partial resets to $x_0$. The reset rate is $r=\lambda/s$, and resets are of the form $x\to x+s(x_0-x)$. For $s=1$ this reduces to a full reset to position $x_0$.

\medskip 

Defining $y\equiv x-x_0$ this translates into a random walk for $y$ subject to resets to $y=0$ with rate $\lambda/s$. We are interested in the mean time until the first arrival at $x=0$, i.e., at $y=-x_0$. In the diffusion approximation, the mean first passage time $T(y_0)$ starting from initial position $y=y_0$ fulfills [c.f. Eq.~(\ref{eq:T_backward})]
\be\label{eq:T_backward2}
(D+\lambda s y^2/2)T''(y)-\lambda yT'(y)+1=0,
\ee
with boundary conditions $T(y=-x_0)=0$ and $T(\infty)=\infty$ (due to the symmetry of the problem we can restrict the analysis to $x>0$, i.e., $y>-x_0$). The solution to this problem can be found in closed form and is given by
\be
T(y)=\int_{-x_0}^y \left(D+\frac{\lambda s}{2}v^2\right)^{1/s}\left(\int_v^\infty \left(D+\frac{\lambda s}{2}u^2\right)^{-1-1/s}du\right)dv.
\ee

The dashed lines in Fig.~2(b) of the main paper are $T(0)$ as obtained from this equation. This is the mean time to reach $y=-x_0$ (equivalent $x=0$) starting from $y=0$ (equivalent $x=x_0$). 

\section{Multiple walkers}
\subsection{Setup}
Now consider multiple walkers with positions $x_i$, $i=1,\dots,N$. We will also refer to the walkers as particles. Between resets the particles move independently according to
\be
\dot x_i(t)= f[x_i(t)]+\xi_i(t),
\ee
where the $\xi_i(t)$ are Gaussian white noise (mean zero) with
\be
\avg{\xi_i(t)\xi_j(t')}=2D\delta_{ij}\delta(t-t').
\ee
In particular $\xi_i$ and $\xi_j$ are independent from one another for $i\neq j$.

Resets occur as before with rate $\lambda/s$. Importantly, all particles are (partially) reset simultaneously when a reset is triggered, i.e.,
\be
x_i \to x_i-sg(x_i)~\forall~i.
\ee

\subsection{Linear system: particle-particle correlations}
\begin{figure}[t]
    \centering
    \includegraphics[width=1\linewidth]{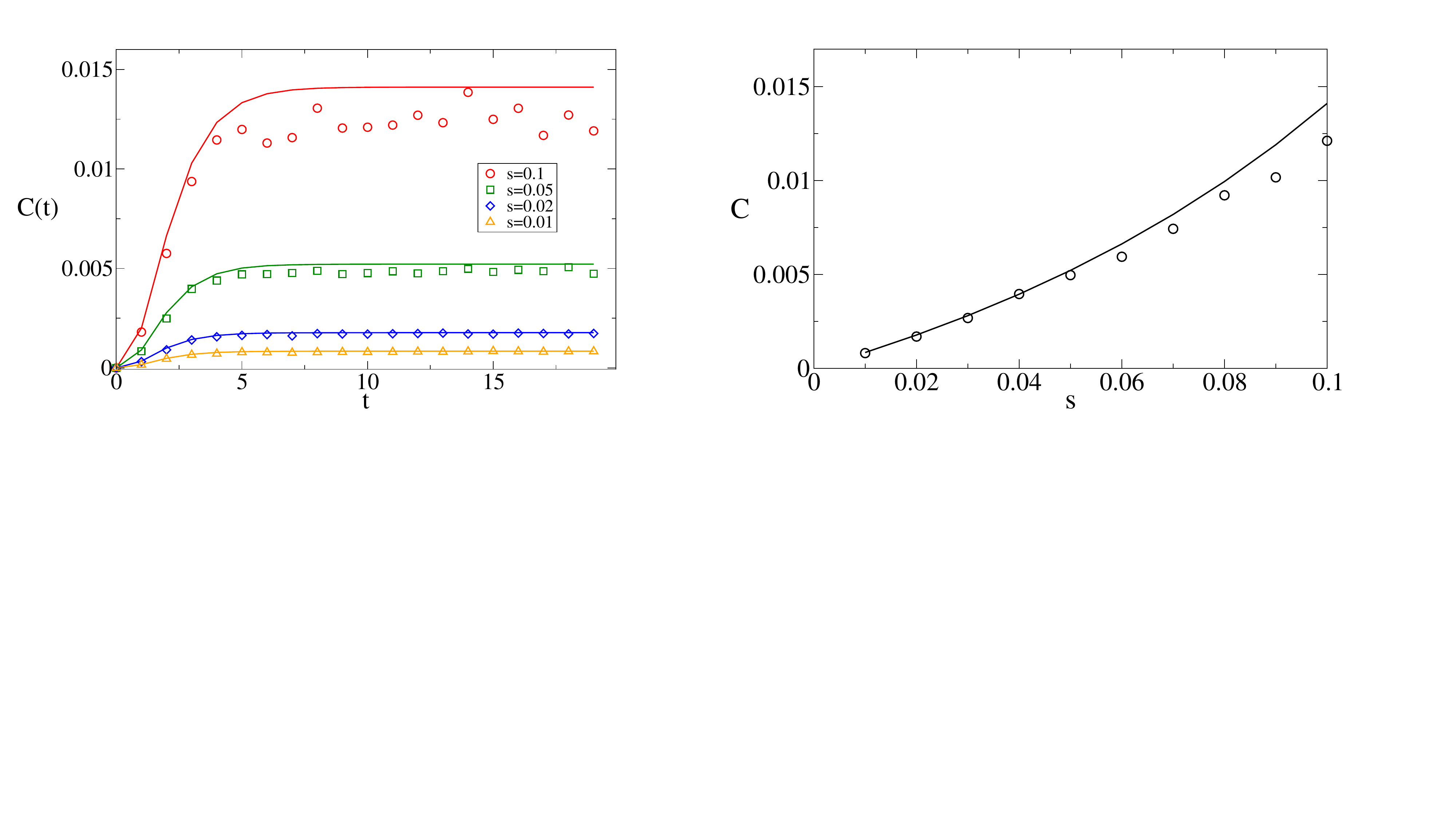}
    \vspace{-15em}
        \caption{Left: Correlator $C(t)= \avg{x_i(t)^2x_j(t)^2}-\avg{x_i(t)^2}\avg{x_j(t)^2}$ for $i\neq j$. Solid lines are from Eq.~(\ref{eq:cxxxx}), markers from simulations (systems with at least $N=100$ particles, depending on the value of $s$, and averaged over $200$ or more realisations). Right: Long-term limit $C(t=\infty)$ as a function of $s$. Solid line is from Eq.~(\ref{eq:correl}), markers are from simulations. Parameters are: $\alpha=0, \gamma=0.5, D=0.1$. Simulations for $N=1000$ particles in panel (a) and $N=200$ particles in (b), results in (b) averaged over time in interval from $10$ to $20$. Data are averages over $5000$ realisations in (a), and $1000$ realisations in (b); integration time step $0.0001$. \label{fig:xxxx}}
\end{figure}

We use again $f(x)=\alpha  + \gamma x$, and additionally we set $\alpha=0$ for simplicity (but the theory can be generalised). We then have
\be\dot x_i(t) = -\mu x_i(t) +\xi_i(t)+\sqrt{\lambda s}\, x_i \eta(t),
\ee
with $\mu\equiv \lambda-\gamma$. From this, we find [assuming $x_i(t=0)=0$ for all $i$]
\be 
x_i(t)=\int_0^t \exp\left[\left(-\mu-\frac{1}{2}\lambda s\right)(t-t')+\sqrt{\lambda s}\left[\Gamma(t)-\Gamma(t')\right] \right]\xi_i(t') dt',
\ee
where $\Gamma(t)=\int^t dt' \eta(t')$. This means that
\be
\avg{x_i(t)|\eta}=0, ~~ \avg{x_i(t) x_j(t')|\eta}=0~(\forall i\neq j),
\ee 
where the notation $\avg{\cdots|\eta}$ means an average conditioned on a fixed realisation of the noise trajectory $\eta$.
We also have
\BE
\avg{x_i(t)^2|\eta}&=& 2D\int_0^t dt' \exp\left[\left(-2\mu-\lambda s\right)(t-t')+2\sqrt{\lambda s}\left[\Gamma(t)-\Gamma(t')\right] \right].
\EE
Next we use  
\be
\avg{\exp\left[2\sqrt{\lambda s}\left[\Gamma(t)-\Gamma(t')\right]\right]}_\eta=\exp\left[2\lambda s (t-t')\right],
\ee
where $\avg{\cdots}_\eta$ stands for an average over noise trajectories $\eta$.
Thus
\BE\label{eq:xxt}
\avg{x_i(t)^2}
&=&\frac{2D}{\lambda s - 2\mu}\left\{\exp\left[\left(-2\mu+\lambda s\right)t\right]-1\right\}.
\EE
At long times, we have
\be
\lim_{t\to\infty} \avg{x_i(t)^2} = \frac{2D}{2\mu-\lambda s}.
\ee
\medskip
For $i\neq j$ we also have
\BE
\avg{x_i(t)^2 x_j(t)^2|\eta}
&=& 2 (2D)^2\int_0^t dt' \int_0^{t'} dt'' \exp\left[\left(-2\mu-\lambda s\right)(2t-t'-t'')\right] \nonumber \\
&&~~~~~~~~~~~~~~~~~~~~~~~~~\times  \exp\left[4\sqrt{\lambda s}\left[\Gamma(t)-\Gamma(t')\right] \right]\exp\left[2\sqrt{\lambda s}\left[\Gamma(t')-\Gamma(t'')\right] \right] \nonumber \\
\EE
In the integral the time $t$ is larger than $t'$, and $t'$ is larger than $t''$. This means that the intervals $[t',t]$ and $[t'',t']$ do not overlap. Therefore, the average over $\eta$ can be carried out as follows
\BE
&&\avg{\exp\left[4\sqrt{\lambda s}\left[\Gamma(t)-\Gamma(t')\right] \right]\exp\left[2\sqrt{\lambda s}\left[\Gamma(t')-\Gamma(t'')\right] \right]}_\eta 
=\exp\left[8\lambda s (t-t')+2\lambda s (t'-t'')\right]
\EE
Thus, we have
\BE\label{eq:xxxxt}
\avg{x_i(t)^2 x_j(t)^2}
&=& 2 (2D)^2\int_0^t dt' \int_0^{t'} dt'' \exp\left[\left(-2\mu-\lambda s\right)(2t-t'-t'')\right] \nonumber \\
&&~~~~~~~~~~~~~~~~~~~~~~~~~\times  \exp\left[8\lambda s (t-t')+2\lambda s (t'-t'')\right] \nonumber \\
&=& 2 (2D)^2\int_0^t dt' \int_0^{t'} dt'' \exp\left[at+bt'+ct''\right] \nonumber \\
&=& 8 D^2 \frac{ce^{at}-(b+c) e^{(a+b)t}+b}{bc(b+c)},
\EE
where I have written
\BE
a&=& -4\mu+6\lambda s, \nonumber \\
b&=&2\mu-5\lambda s, \nonumber \\
c&=&2\mu-\lambda s,
\EE
and where we have used $a+b+c=0$. We note that $a<0$ and $a+b<0$ for small enough $s$. The long-time limit is therefore
\be
\lim_{t\to\infty} \avg{x_i(t)^2 x_j(t)^2}=\frac{8D^2}{(2\mu-\lambda s)(4\mu-6\lambda s)}.
\ee
\medskip
Using Eqs.~(\ref{eq:xxt}) and (\ref{eq:xxxxt}) we can compute the correlator
\be\label{eq:cxxxx}
C(t)\equiv \avg{x_i(t)^2x_j(t)^2}-\avg{x_i(t)^2}\avg{x_j(t)^2}
\ee
for $i\neq j$. Results are confirmed in simulations, as shown in Fig.~\ref{fig:xxxx}.
\medskip

 The long-term limit of this quantity is
\BE\label{eq:correl}
\lim_{t\to\infty}\left[ \avg{x_i(t)^2 x_j(t)^2} - \avg{x_i(t)^2} \avg{x_j(t)^2}\right]&=&\frac{8D^2}{(2\mu-\lambda s)(4\mu-6\lambda s)} - \frac{4D^2}{(2\mu-\lambda s)^2} \nonumber \\
&=& \frac{8D^2}{2\mu-\lambda s}\left[\frac{1}{4\mu-6\lambda s}-\frac{1}{4\mu-2\lambda s}\right].
\EE
We recall that the diffusion approximation that was used to derive this expression is valid for small $s$. For $\lambda>0$ and $s$ small but positive, the expression in Eq.~(\ref{eq:correl}) is positive, confirming the presence of dynamically induced correlations. For $\lambda=0$ or $s=0$, the expression vanishes (to consider this limit we assume that $\gamma>0$ so that there is reversion to the origin in the process without resetting). This limit describes a system without resets, hence there is no common dynamics  which could generate correlations. The data in Fig.~\ref{fig:xxxx} indicates that these correlations increase with $s$ in our system, i.e., as resets become less frequent but have a stronger effect.

\section{Individual-based model}
\subsection{Model definition and diffusion approximation without catastrophes}
We consider a population of discrete individuals, which we write as $A$. The number of individuals in the population is $n$. This is a time-dependent random variable. The population size changes via birth and death events, which we write as 
\BE
A&\stackrel{1}{\rightarrow}& A+A ,\nonumber \\
A+A&\stackrel{1}{\rightarrow}& A.
\EE
The first reaction describes proliferation events, in which one individual gives birth to another. We assume that this occurs with constant unit per capita rate per individual, so that the rate of reactions of this type in the population is $T_1(n)=n$ (if there are $n$ individuals in the population, the expected number of reproduction events is $n$ per unit time). The second reaction captures competition and leads to a reduction of the population. In the population reactions of this type are taken to occur with rate $T_2(n)=n^2/\Omega$, where $\Omega$ is a parameter setting the scale of the population size. 

In absence of resetting, and taking a deterministic limit, this population is described by the rate equation $\frac{d}{dt}\avg{n} = \avg{n}-\avg{n}^2/\Omega$. This is a logistic growth equation with non-trivial fixed point $\avg{n}=\Omega$. It is in this sense that $\Omega$ sets the scale for the population size.

The master equation for these dynamics is
\be
\frac{d}{dt} P(n,t)=(n-1)P(n-1,t)+\frac{(n+1)^2}{\Omega} P(n+1,t)-\left(n+\frac{n^2}{\Omega}\right)P(n,t).
\ee

We now write $x=n/\Omega$. We can now follow standard steps \cite{gardiner2004handbook, van1992stochastic} and carry out a Kramers--Moyal expansion of the master equation for the probability distribution $P(n,t)$. This is not to be confused with the expansion for small $s$, at this point we have not yet introduced catastophes (resets). Instead the expansion is valid for large but finite populations ($\Omega \gg 1$). One then finds,
\be
\dot x = x(1-x)+\sqrt{\frac{x+x^2}{\Omega}}\xi.
\ee
The noise $\xi$ is Gaussian, of mean zero and with $\avg{\xi(t)\xi(t')}=\delta(t-t')$. It captures {\em intrinsic} demographic fluctuations in the population, due to the random occurrence of the birth and death events.
\subsection{Additional catastrophes}
We also allow for random catastrophes (resets) occurring with rate $r=\lambda/s$. In each reset event any existing individual independently dies with probability $s$, or survives with probability $1-s$. This models binomial catastrophes.

In a catastrophe, we have $\avg{\Delta n}=-sn$, and $\mbox{Var}(\Delta n)=ns(1-s)\approx ns$ for $s\ll 1$. This means (using $x=n/\Omega$) that $\avg{\Delta x}=-sx$ and $\mbox{Var}(\Delta x)=xs/\Omega$. Thus, we can write
\be
\Delta x = -sx(1+z),
\ee
with Gaussian random variable $z$, with zero average and $\mbox{Var}\,z=\frac{1}{\Omega sx}$. Thus, using the results in Sec.~A3 in the End Matters in the main paper we have, 
\be\label{eq:aux1}
\dot x = x(1-x)+\sqrt{\frac{x+x^2}{\Omega}}\xi-\lambda x +\sqrt{\lambda s x^2+\frac{\lambda}{\Omega}x}\,\eta,
\ee
with $\avg{\eta(t)\eta(t')}=\delta(t-t')$. 
\medskip

The first term under the root multiplying $\eta$ in Eq.~(\ref{eq:aux1}) captures fluctuations in the number of resets that occur per unit time. This is as in the previous sections. The second term, $(\lambda/\Omega) x$, represents the varying size of the catastrophes. These are binomial in the full model, but are approximated as Gaussian in the diffusion approximation. For infinite populations ($\Omega\to\infty$) the binomial becomes a delta function, and hence there are no fluctuations of the number of individuals that die in a reset. The contribution reflecting the variation of catastrophe sizes then vanishes.

\subsection{Linear-noise approximation}
Ignoring the noise terms in Eq.~(\ref{eq:aux1}) we find the non-zero deterministic fixed point $x^\star=1-\lambda$. Writing $x=x^\star+y$, and expanding to linear order in $y$, we further have, making the linear-noise approximation,
\be\label{eq:lna_ou}
\dot y= (1-2x^\star-\lambda) y + \sigma\zeta,
\ee
with $\zeta$ Gaussian white noise of zero mean and unit variance, and where $\sigma^2=\frac{x^\star+(x^\star)^2}{\Omega}+\lambda s (x^\star)^2+\frac{\lambda}{\Omega}x^\star$.

\medskip

Eq.~(\ref{eq:lna_ou}) is an Ornstein-Uhlenbeck process, and we note that $1-2x^\star-\lambda=-1+\lambda$ is negative provided $\lambda<1$ which we will always assume (this condition is also needed to ensure that $x^\star=1-\lambda$ is positive). The stationary distribution of this process is Gaussian, centred on $y=0$ and with variance $\Sigma^2=\sigma^2/[-2(1-\lambda-2x^\star)]$. Thus, the distribution of $x$ is centred on $x^*$ and has the same variance $\Sigma^2$. From this, the distribution of $n=x\Omega$ can be obtained by re-scaling. This is the distribution shown as a solid line in the lower panel of Fig.~3 in the main paper.

\section{Resetting-induced cycles and patterns}
\subsection{Resetting induced quasi-cyles in a predator-prey model}
We start from the predator-prey (see e.g. \cite{mckane_newman})
\BE
\dot x &=&xy-ax, \nonumber \\
\dot y&=&by(1-y/k)-cxy,
\EE
where $x$ represents the abundance of predators, and $y$ that of prey. The terms proportional to $xy$ reflect predation, and the term $-ax$ indicates attenuation of the predator population (if there is no prey). The prey population, finally, follows logistic growth in absence of predators, as indicated by the term $by(1-y/k)$.

We again assume random catastrophes, occurring as  Poisson process with rate $\lambda/s$. Resetting only affects the prey species $y$ in our setup. If a catastrophe occurs, then $y\to (1-s) y$. 

\medskip

Using the theory in the End Matters in the main paper, we find the following diffusion approximation for small, but non-zero $s$,
\BE\label{eq:pp_diff_approx}
\dot x &=&xy-ax, \nonumber \\
\dot y&=&by(1-y/k)-cxy-\lambda y +\sqrt{\lambda s y^2} \eta,
\EE
with $\avg{\eta(t)\eta(t')}=\delta(t-t')$. We write $x^\star$ and $y^\star$ for the components of the fixed point of Eqs.~(\ref{eq:pp_diff_approx}) in the deterministic limit (i.e., with the noise term $\eta$ removed).

\medskip

Linearising about this fixed point (where we write $x=x^\star+u$, $y=y^\star+v$), and making a linear-noise approximation we then find
\BE\label{eq:LNA1}
\dot u &=& (y^\star-a) u + x^\star v, \nonumber \\
\dot v &=& - c y^\star u + [b(1-2y^\star/k) -c x^\star-\lambda ] v   +\sqrt{\lambda s} y^\star\zeta,
\EE
with $\avg{\zeta(t)\zeta(t')}=\delta(t-t')$.
This can be written in the form
\BE
\dot u &=& J_{11} u + J_{12} v, \nonumber \\
\dot v &=& J_{21} u + J_{22} v+\sqrt{B_{22}}\zeta,
\EE
where the entries $J_{ij}$ of the Jacobian can be read off from Eq.~(\ref{eq:LNA1}), and where we introduce $B_{22}=\lambda s (y^\star)^2$. Following \cite{mckane_newman} one then carries out a Fourier transform (indicated by tildes below). After some algebra one finds
\be\label{eq:pp_spectrum}
S_i(\omega)=\frac{\alpha_i+\beta_i\omega^2}{(\omega^2-\Omega_0^2)^2+\Gamma^2\omega^2},
\ee
where we define $S_1(\omega)\equiv \avg{|\tilde u(\omega)|^2}$ and $S_2(\omega)\equiv \avg{|\tilde v(\omega)|^2}$, and where we have for our system,
\BE
\alpha_1&=&B_{22}J_{12}^2, \nonumber \\
\alpha_2&=&B_{22}J_{11}^2, \nonumber \\
\beta_1&=&0, \nonumber \\
\beta_2&=&B_{22},
\EE
 and
\BE
\Omega_0^2=\det \underline{\underline{J}}, \nonumber \\
\Gamma^2=(\mbox{tr} \underline{\underline{J}})^2.
\EE
The theory lines in Fig.~4 of the main paper are from Eq.~(\ref{eq:pp_spectrum}).
\subsection{Resetting-induced quasi patterns}
We consider the Levin--Segel model of plankton-herbivore dynamics \cite{levin1976hypothesis}. We write $\psi(\bx,t)$ for the concentration of plankton at position $\bx$ and time $t$, and $\phi(\bx,t)$ for the concentration of the herbivore species. Using the notation of \cite{butlergoldenfeld}, the model is governed by partial differential equations
\BE\label{eq:LSmodel}
\frac{\partial}{\partial t} \psi&=&D_\psi \Delta \psi + b\psi+e\psi^2-(p_1+p_2)\psi\phi, \nonumber \\
\frac{\partial}{\partial t}\phi&=&D_\phi \Delta\phi+p_2\phi\psi-d\phi^2.
\EE
Again following \cite{butlergoldenfeld} we set $p_1=0, p_2=1$ and we fix $b=d=e=1/2$. 

\begin{figure}[t!!!]
\vspace{1em}
   \begin{center}
    \includegraphics[width=0.8\linewidth]{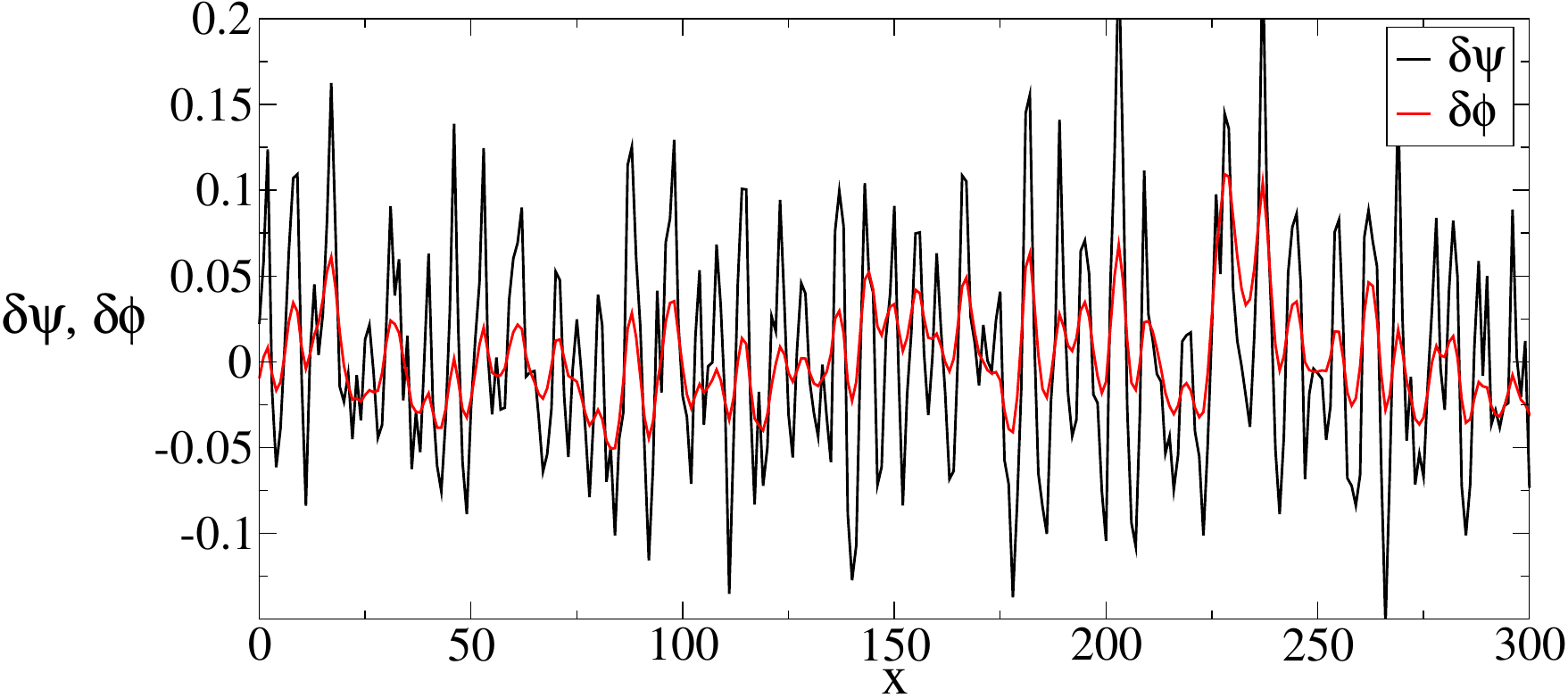} 
    \end{center}
    \caption{Resetting induced pattern in the Levin--Segel model with resets in one spatial dimension. The figure shows a single realisation. Model parameters: $b=d=e=1/2$, $p=1$, $D_\psi=0.1$, $D_\phi=2.5$, $\lambda=0.1$, $s=0.05$. Spatial discretisation is in steps of $\Delta x=1$, spatial extent is $2048$ units, with periodic boundary condition. Time steps for the Euler-forward integration are $\Delta t=0.001$. 
    \label{fig:1dpattern}}
\end{figure}

We allow for random catastrophes acting as proportional resets on the herbivore species. Resets occur with rate $\lambda/s$, as an independent Poisson process at each location in space. When such an event occurs the local herbivore density $\psi(\bx,t)$ is reset as $\psi(\bx,t)\to (1-s)\psi(\bx,t)$.
\medskip

Implementing the diffusion approximation for $s\ll 1$, this leads to the stochastic partial differential equations
\BE\label{eq:diff_approx_PDE}
\frac{\partial}{\partial t} \psi&=&D_\psi \Delta \psi + \frac{1}{2}\psi+\frac{1}{2}\psi^2-\psi\phi-\lambda \psi +\sqrt{\lambda s}\psi\eta, \nonumber \\
\frac{\partial}{\partial t}\phi&=&D_\phi \Delta\phi+\psi\phi-\frac{1}{2}\phi^2,
\EE
with $\avg{\eta(t)\eta(t')}=\delta(t-t')$.

\medskip

In the limit $s\to 0$ the stochastic partial differential equation (\ref{eq:diff_approx_PDE}) becomes deterministic. The spatially uniform fixed-point solution of the resulting deterministic equations is 
\be
\psi^*=\frac{2}{3}\left(\frac{1}{2}-\lambda\right),~~~
\phi^*=\frac{4}{3}\left(\frac{1}{2}-\lambda\right).
\ee
The Jacobian at this fixed point has the following entries (where we write $i=1$ for $\psi$, and $i=2$ for $\phi$):
\BE
J_{\bk,11}&=&-D_\psi \bk^2+\frac{1}{3}\left(\frac{1}{2}-\lambda\right), \nonumber \\
J_{\bk,12}&=&-\frac{2}{3}\left(\frac{1}{2}-\lambda\right), \nonumber \\
J_{\bk,21}&=&\frac{4}{3}\left(\frac{1}{2}-\lambda\right),\nonumber \\
J_{\bk,22}&=&-D_\phi \bk^2-\frac{2}{3}\left(\frac{1}{2}-\lambda\right).
\EE
We have taken the Fourier transform in the diffusion terms, writing $\bk$ for the corresponding wave vector. The above form, involving factors of $-\bk^2$, applies in a continuous space. The spectrum shown in Fig. 5 of the main paper is from a simulation one a one-dimensional discrete lattice ($dx=1$), corresponding to a meta-population model with discrete patches. The factor $-\bk^2$ in the Jacobian is then to be replaced by $2\left[\cos(k)-1\right]$.

\medskip

From the Jacobian a condition for the onset of Turing patterns can be derived, see Eq. (4) in \cite{butler2009robust} (this is for general parameters $D_\psi, D_\phi, b, d, e, p_1, p_2$). We note that the parameter $b$ does not feature in this condition. We also note that the term $-\lambda\psi$ in the first equation in (\ref{eq:diff_approx_PDE}) can be absorbed in the term $b\psi$ by a redefinition of $b$. Therefore, the term $-\lambda\psi$ (which is due to the resetting, and not present in the model in \cite{butler2009robust, butlergoldenfeld}), does not affect the onset of Turing patterns in the deterministic system $(s=0)$. For our choice of parameters  the deterministic system shows Turing patterns when $D_\phi/D_\psi>27.8$  \cite{butler2009robust, butlergoldenfeld}. We will generally studying the model below this threshold, so that there are no Turing patterns in the absence of noise.

\medskip

Restoring the noise term and carrying out a linear-noise approximation, combined with a linear expansion about the fixed point, we have (writing $\delta\psi$ and $\delta\phi$ for deviations from the fixed point)
\BE
\delta\psi(\bk,\omega)&=& J_{\bk,11}\delta\psi(\bk,\omega)+J_{\bk,12}\delta\phi(\bk,\omega)+\zeta_1(\bk,\omega) \nonumber \\
\delta\phi(\bk,\omega)&=& J_{\bk,21}\delta\psi(\bk,\omega)+J_{\bk,22}\delta\phi(\bk,\omega)+\zeta_2(\bk,\omega),
\EE
where the $\zeta_i(\bk,\omega)$ are Gaussian random variables. We write the correlation matrix as $\avg{\zeta_i(\bk,\omega)\zeta_j(\bk,\omega)}=B_{\bk,i,j}\delta(\omega-\omega')\delta(\bk-\bk')$, and then have
\BE
B_{k,11}=\lambda s (\psi^*)^2,~~~B_{k,12}=B_{k,21}=B_{k,22}=0.
\EE

Following \cite{lugo2008quasicycles, butlergoldenfeld} the power spectrum of fluctuations is then given by 
\BE
P_1(\bk,\omega)&=&\frac{C_{k,1}+B_{\bk,11}\omega^2}{(\omega^2-\Omega_{0,\bk}^2)^2+\Gamma_\bk^2\omega^2} \nonumber \\
P_2(\bk,\omega)&=&\frac{C_{\bk,2}+B_{\bk,22}\omega^2}{(\omega^2-\Omega_{0,\bk}^2)^2+\Gamma_\bk^2\omega^2}.
\EE
with 
\BE
\Omega_{0,\bk}^2&=&\det\,\underline{\underline{J}}_\bk, \nonumber \\
\Gamma_k^2&=&(\mbox{tr}\, \underline{\underline{J}}_\bk)^2.
\EE
and where for our model 
\BE
C_{\bk,1}&=& B_{\bk,11}J_{\bk,22}^2, \nonumber \\
C_{\bk,2}&=&B_{\bk,11}J_{\bk,21}^2.
\EE

An example of a resetting-induced pattern in one dimension is shown in Fig.~\ref{fig:1dpattern}. 

The power spectrum $P_1(k,\omega=0)$ in $k$-space for the one-dimensional model shown in Fig.~5 in the main paper was obtained from an average over $300$ realisations, with model parameters as in the caption of Fig.~\ref{fig:1dpattern}.